\documentclass[aps,preprint,showpacs,preprintnumbers,amsmath,amssymb]{revtex4}
\usepackage[dvips]{graphicx}
\usepackage{float}
\begin{document}

\title{The de Sitter QED in Coulomb gauge: \\
first order transition amplitudes}
\author{Ion I. Cot\u aescu \thanks{E-mail:~~cota@physics.uvt.ro}
and Cosmin Crucean \thanks{E-mail:~~crucean@physics.uvt.ro}\\
{\small \it West University of Timi\c soara,}\\
{\small \it V. Parvan Ave. 4 RO-300223 Timi\c soara,  Romania}}

\begin{abstract}
We construct the de Sitter QED in Coulomb gauge assuming that the quantum modes are prepared by a global apparatus which is able to determine a stable and invariant vacuum state, independent on the local coordinates. Then we proceed in traditional manner postulating the appropriate equal-time commutators and anti-commutators of the interacting fields and deriving the perturbation expansion of the scattering operator. In this approach the $in -out$ transitions amplitudes, measured by the same global apparatus, can be calculated exactly by using the reduction formalism and the perturbation procedure as in the flat case but with significant differences due to the de Sitter geometry.  A specific feature is that the gravity eliminates the constraints due to the simultaneous momentum-energy conservation giving rise to QED transitions with non-vanishing amplitudes even in the first order of perturbations.  Of a special interest could be the first order amplitudes of the electromagnetic particle creation allowed by the expansion of the de Sitter universe. We show that this effect is significant only in the very strong gravity of the early universe.
\end{abstract}


\maketitle


\section{Introduction}

The theory of quantum fields  is governed by the fundamental principles of quantum physics which say that the quantum states are prepared  by a classical apparatus that measures quantum observables (i. e. operators acting on the state space). In special relativity the Poincar\' e isometries give rise to a set of conserved observables globally defined that enable one to define bases of the state space formed by the common eigenstates of some complete sets of commuting operators. In this framework one can appeal to a natural prescription for separating  invariant subspaces of particles and antiparticles corresponding to a stable vacuum state.

In general relativity the curved spacetimes could have no isometries and no conserved observables. Then the quantum modes are often defined locally by choosing the convenient solutions of the field equations in each local chart  (or
natural frame) separately. Thus the vacuum could become unstable in the sense that the quantum modes of two different local charts  may be related through Bogoliubov transformations which mix the particle and antiparticle modes among themselves \cite{BD}. This is the mechanism of the cosmological particle creation giving rise to local effects in  accelerated frames as, for example, the Unruh \cite{U,HH} and Gibbons-Hawking \cite{GH} ones. These may be observed using the {local} particle detectors proposed by Urunh \cite{U} and DeWitt \cite{DW}.

However, there are special cases when the curved local-Minkowskian manifolds have isometries that could take over the role of the Poincar\'e symmetry. The theory of quantum fields with spin on such manifolds  can be correctly constructed only in orthogonal (non-holonomic) local frames where the half-integer spins do make sense \cite{WALD,SG}. We have shown that this is a tetrad-gauge covariant Lagrangian theory in which  the matter fields transform under isometries according to the {\em covariant} representations (CR) {\em induced}  by the (non-unitary) finite-dimensional ones of the $SL(2,\Bbb C)$ group which is the universal covering group of the gauge group (of the Minkowski metric), $L^{\uparrow}_{+}\subset SO(1,3)$ \cite{ES}.  The generators of the CRs are the differential operators produced by the Killing vectors associated to isometries according to the  generalized Carter and McLenagan formula \cite{CML,ES,EPL}. They form an algebra of conserved observables (commuting with the operators of the field equations) among them one can select the complete sets of commuting operators able to define different bases of the state space. With this method the quantum states are {\em globally} defined on the entire manifold, independent on the local coordinates, so that we can find a suitable prescription providing us with a unique and stable vacuum state. This  conjecture is closer to the genuine principles of quantum mechanics since the quantum states are {\em prepared} and measured by the {\em same}  global apparatus represented by the largest freely generated algebra that includes the conserved operators. However, this does not exclude or contradict the cosmological particle creation which may be observed  by using  local detectors that work in a different manner \cite{HH}.

Our approach is helpful on the de Sitter  space-time where all the free field equations can be analytically solved, while the specific  $SO(1,4)$ isometries offer us  conserved observables with well-defined physical meaning \cite{CCC}  that form a $so(1,4)$ Lie algebra.  However, a special problem of this geometry is that the only time-like Killing vector is not time-like everywhere. This generated some doubts concerning the possibility of defining correctly the energy operator, the $in$ and $out$ fields as well as the scattering operator \cite{Witt}. Nevertheless, we have shown that there are no real impediments since the time-like Killing vector keeps this property everywhere {\em inside} the light-cone where an observer can perform physical measurements \cite{CCC}. Moreover,  in this geometry the light-cone domain of any observer remains behind his horizon such that the measurements can be done as in the flat case without to be affected by the presence of the observer's horizon.  Thus the energy operator is correctly defined,  but it does not commute with the momentum operator and, consequently, the dispersion relation is absent. This affects the measurements of the energy and momentum, which are diagonal in different bases, called here the momentum and respectively energy representations \cite{CD1}-\cite{Max}, \cite{Cs}. Under such circumstances, we derived the  quantum modes of both these bases for the free Dirac \cite{CD1,CD2,CD3}, Proca \cite{Cv} and Maxwell \cite{Max} fields, performing the {\em canonical} quantization with a stable and invariant vacuum state.

We specify that the Dirac quantum modes correctly normalized in momentum representation were found many years ago by Nachtmann \cite{Nach}, which
considered another method of constructing CRs that, in the de Sitter case, is equivalent to our theory of external symmetry \cite{ES} up to some conventions and notations. Moreover, Nachtmann indicated how the covariant scalar and Dirac fields can be quantized in canonical manner. As a matter of fact, we continue his line attempting to construct a coherent quantum field theory on the de Sitter space-time \cite{CD1}-\cite{Max}, even though sometime our results differ from those found by other authors. This is because our induced CRs are less used in literature, since one  prefers either to construct directly maximally symmetric two-point functions \cite{Wood}, or to use only linear representations \cite{Gaz}.

According to the above arguments, we believe that our approach is appropriate for investigating quantum effects on the de Sitter expanding universe, including the catching mechanisms of the particle creation. The idea of the space expansion generating quantum matter is very old \cite{Schr} but the technical difficulties and the contradictory results obtained so far discourage definitive conclusions. This effect was studied first in models with free (non-interacting) quantum fields coupled to gravity. A functional model exactly solved by  Candelas and Raine leads to the conclusion that the whole de Sitter manifold can not create particles \cite{CanRe}. On the contrary, using a WKB-type method Parker found that this phenomenon does exist in various expanding Robertson-Walker universes \cite{WKB}. Other authors considered the most realistic conjecture of the QED on  such expanding space-times \cite{Lot} or only on the de Sitter one \cite{rusi,IITsar}. A general method of investigating QED effects in the first order of perturbations was proposed by Lotze, which calculates the $in-out$ transition probabilities \cite{Lot} taking into account the 'added-up' decay probability prescribed by Audretsch and Spangehl \cite{AuSp}. In this approach the new effects of pair creation and electromagnetic particle creation from the QED vacuum were analysed. Unfortunately, this method is too general for giving concrete results without approximations and the problem of particle production in the early universe is not addressed. Moreover, there are no exact results even in the de Sitter QED where the electromagnetic particle creation  was studied but in a wrong geometric context, adding the opposite contributions of the expanding and collapsing portions \cite{IITsar}. The effect of fields interactions (scalar and electromagnetic fields) upon particle creation was studied in \cite{BR} using a S matrix approach that allows one to estimate the number of particles created in this context.

In the present paper we would like  to continue this line, restricting ourselves to the de Sitter QED in Coulomb gauge. We study the effect of the electromagnetic interaction upon the particle creation in the de Sitter expanding universe. We  prove that this kind of particle production arise only in the early universe, when the expansion factor is larger than the particle mass. This result confirms  the Parker's one \cite{WKB} which establish that the rate of pair production in purely gravitational field was important only in the early universe.

We start our study assuming that the QED transitions are measured by the {\em same} global apparatus which prepares all the quantum states, including  the $in$ and $out$ asymptotic free fields which remain  minimally coupled to gravity, but without electromagnetic interaction. This apparatus complies with an asymptotic prescription of frequencies separation that assures the uniqueness and stability of both the vacuum states of the free Dirac and Maxwell fields \cite{CD1,CD3,Max}. In other words it can  not record particle creation in the absence of the electromagnetic interaction, as the local detectors can do. This context inhibits the cosmological particle creation allowing us to concentrate only on the QED processes in order to obtain {\em exact} results using perturbations as in special relativity \cite{BDR,SW1}. To this end we construct the de Sitter QED in Coulomb gauge starting with the Lagrangian theory that gives the field equations and the principal conserved observables of the interacting fields produced by the $SO(1,4)$ symmetry via Noether's theorem. Furthermore, we postulate the equal-time commutators and anti-commutators  and  derive the equation of the time-dependent evolution operator. Hereby we obtain the perturbation series of the scattering operator in terms of free fields. This is generated by a specific interaction Hamiltonian which does not depend on the Coulomb potential. We recover thus the  framework of the QED in special relativity where the Coulomb gauge allows a natural quantization separating in fact the Coulomb potential. Finally, we define the asymptotic fields showing how the $in-out$ amplitudes can  be calculated by using  the  reduction formalism and the mentioned scattering operator.

As examples, we study  all the simple QED processes which have non-vanishing $in-out$ amplitudes in the {\em first order} of perturbations. These are of interest  since correspond to transitions which in the flat limit are forbidden by the energy-momentum conservation \cite{Lot}. We specify that our calculations are exact, without any kind of approximation, so that our results hold for any values of parameters allowing us to consider strong gravitational fields or extremely large (or small) masses, energies and momenta. We can thus analyse the behaviour of the transition probabilities of the following effects:  electrons or positrons absorbing or emitting one photon, one photon pair creation or annihilation and the particle creation or annihilation from or into the QED vacuum.

The paper is organized as follows. In the second section  we briefly present our previous results concerning  the canonical quantization of the free Dirac and Maxwell fields  considering only the momentum bases in  the (co)moving chart of conformal time and the diagonal tetrad-gauge \cite{CD1,Max}. The next section is devoted to the de Sitter QED in Coulomb gauge. We give the specific equal time algebraic rules, define the asymptotic fields and present the reduction method and the perturbation expansion of the scattering operator.  In section 4 the first order $in-out$ transition amplitudes are calculated in terms of hypergeometric functions. In section 5 we give and comment our principal numerical results concerning the behaviour of the corresponding probabilities. Finally, we present our concluding remarks.

\section{Free fields on the de Sitter space-time}

Let $(M_{\omega},g)$ be the de Sitter space-time of Hubble constant $\omega$.  The covariant fields on this manifold are defined in local charts of coordinates $x^{\mu}$, ($\mu,\nu,...=0,1,2,3$) and orthogonal (non-holonomic) local frames determined  by tetrad fields, $e_{\hat\mu}$  and $\hat e^{\hat\mu}$. These are labelled by local indices ($\hat\alpha,\hat\beta,...=0,1,2,3$) which are raised or lowered by the Minkowski metric $\eta={\rm diag}(1,-1,-1,-1)$ while for the natural indices we have to use the metric tensor $g_{\mu \nu}=\eta_{\hat\alpha\hat\beta}\hat
e^{\hat\alpha}_{\mu}\hat e^{\hat\beta}_{\nu}$.

Here we consider the moving chart $\{t,\vec{x}\}$ of {\em conformal} time, $t\in (-\infty,0)$, Cartesian coordinates and line element
\begin{equation}
ds^{2} =\frac{1}{(\omega t)^2}\,\left(dt^{2}- d\vec{x}\cdot d\vec{x}\right)\,,
\end{equation}
which covers the expanding portion of the de Sitter manifold. In this chart we fix the {\em diagonal} tetrad-gauge,
\begin{equation}\label{tt}
e^{0}_{0}=-\omega t\,, \quad e^{i}_{j}=-\delta^{i}_{j}\,\omega t \,,\quad \hat
e^{0}_{0}=-\frac{1}{\omega t}\,, \quad \hat e^{i}_{j}=-\delta^{i}_{j}\,
\frac{1}{\omega t}\,,
\end{equation}
defining the local orthogonal frame $\{e\}$ and the corresponding co-frame
$\{\hat e\}$. Then we have  $\sqrt{g}=\sqrt{|{\rm det}(g)|}={\rm det} (\hat e)=(\omega t)^{-4}$. We remind the reader that the {\em proper} time $\hat t=-\frac{1}{\omega}\ln(-\omega t)\in (-\infty,\infty)$ of the moving chart $\{\hat t,\vec{x}\}$ is that which tends to the physical one in the flat limit when $\omega\to 0$ and $(-\omega t)\to 1$.

The de Sitter manifold is a homogeneous space of the pseudo-orthogonal group $SO(1,4)$ which plays the role of isometry group. The generators of the CRs given by the corresponding Killing vectors  have orbital and spin terms which, in general,  do not commute among themselves \cite{ES,EPL}. In our frames $\{t,\vec{x}\}$ and $\{e\}$ there are only two operators which do not have spin parts namely, the momentum operator $\vec{P}=-i\vec{\partial}$ (corresponding to the Killing vectors $k_{(i)}$ of components $k_{(i)}^0=0$ and $k^j_{(i)}=\delta^j_i$) and the energy operator,
\begin{equation}\label{hami}
H=-i\omega (t\partial_t+x^i\partial_i)\,,
\end{equation}
(given by the Killing vector $k^{\mu}=\omega x^{\mu}$) \cite{CD1,CCC}. These operators  do not commute to each other, $[H,\vec{P}]=i\omega \vec{P}$, so that they are diagonal in two different bases. These are the momentum basis in which $P_i$ are diagonal and the energy basis where $H$ is diagonal simultaneously with the (non-differential) operators of the momentum direction \cite{Cs,CD2,Max}.

We consider here only the momentum  representation where the free field equations can be analytically solved in terms of Bessel functions. Since the energy operator is not diagonal in this basis it can not be used for separating the positive and negative frequencies modes in any frame. Therefore, we adopt an usual asymptotic prescription assuming that \cite{CD1,Max}: in the chart under consideration {\em the particle mode functions oscillate as progressive waves when $t\to -\infty$}. Hereby we recover the conformal vacuum states of the Maxwell theory \cite{Max} while in the Dirac case we obtain a correct separation of the  particle and antiparticle spinors in the rest frames where they become eigenspinors of the energy operator \cite{CD3}. Thus we define a unique Dirac vacuum state which is just an {\em adiabatic} vacuum, independent on the natural or local frames we chose, as we shall show later. Our previous results and some preliminary calculations indicate that both the vacuum states we use here are {\em invariant} under the $SO(1,4)$ isometries.

\subsection{The free Dirac field}

The free Dirac field $\psi$ of mass $m$ and minimally coupled to the gravity of $M_{\omega}$ has the action
\begin{equation}\label{actionD}
{\cal S}[\psi]=\int d^4 x\sqrt{g}\, {\cal L}_D(\psi)\,.
\end{equation}
where the Lagrangian density,
\cite{CD1},
\begin{equation}
{\cal L}_D(\psi)=
\frac{i}{2}\,[\overline{\psi}\gamma^{\hat\alpha}D_{\hat\alpha}\psi-
(\overline{D_{\hat\alpha}\psi})\gamma^{\hat\alpha}\psi] -
m\overline{\psi}\psi\,,\quad \overline{\psi}=\psi^+\gamma^0\,,
\end{equation}
depends on  the covariant derivatives in local frames, $D_{\hat\alpha}$ \cite{CD1}, that guarantee the tetrad-gauge invariance. The point-independent Dirac matrices $\gamma^{\hat\mu}$ satisfy $\{ \gamma^{\hat\alpha},\, \gamma^{\hat\beta} \}=
2\eta^{\hat\alpha \hat\beta}$ giving rise to the basis-generators
$S^{\hat\alpha \hat\beta}=i [\gamma^{\hat\alpha}, \gamma^{\hat\beta}
]/4$ of the spinor representation $(\frac{1}{2},0)\otimes (0,\frac{1}{2})$ of the $SL(2,\Bbb C)$ group that induces the spinor CR \cite{ES,CD1}. The Lagrangian density remains invariant under the internal $U(1)_{em}$ transformations, $\psi\to e^{-i\alpha}\psi$, while the whole action (\ref{actionD}) is invariant under isometries as long as the Dirac field transforms according to the spinor CR.

In the chart $\{t,\vec{x}\}$ and tetrad-gauge (\ref{tt}), the free Dirac equation \cite{CD1},
\begin{equation}\label{EDir}
E_D(x)\psi(x)=\left[-i\omega t\left(\gamma^0\partial_{t}+\gamma^i\partial_i\right)
+\frac{3i\omega}{2}\gamma^{0}-m\right]\psi(x)=0\,,
\end{equation}
can be put in Hamiltonian form, $H\psi=H_D\psi$, using the energy operator (\ref{hami}) and the Hamiltonian operator
\begin{equation}
H_D=i\omega t \gamma^0\vec{\gamma}\cdot\vec{\partial}+m\gamma^0-i\omega\left(\vec{x}\cdot\vec{\partial}+\frac{3}{2}\right)\,,
\end{equation}
whose last term represents the gravitational energy of the minimal coupling \cite{CSchr}. This equation can be analytically solved either in momentum or energy bases with correct orthonormalization and completeness properties \cite{CD1,CD2} with respect to the relativistic scalar product of the Dirac theory
\begin{equation}
\left< \psi,\psi'\right>_D=\int d^3x\,(-\omega t)^{-3}\overline{\psi}(t,\vec{x})\gamma^0 \psi'(t,\vec{x})\,.
\end{equation}

The mode expansion in momentum representation,
\begin{equation}\label{psiab}
\psi(t,\vec{x})=\int d^3 p \sum_{\sigma}\left[U_{\vec{p},\sigma}(x)
a(\vec{p},\sigma)+V_{\vec{p}, \sigma}(x){a^c}^{\dagger}(\vec{p},
\sigma) \right]\,,
\end{equation}
is written in terms of the field operators, $a$ and $a^c$, and  the particle and antiparticle fundamental spinors of this basis, $U_{\vec{p},\sigma}$ and respectively $V_{\vec{p},\sigma}$,  which depend on the momentum $\vec{p}$ (with $p=|\vec{p}|$) and polarization $\sigma=\pm 1/2$. According to our prescription of frequencies separation we find that these spinors,  in the standard representation of the Dirac matrices (with diagonal $\gamma^0$), take the form \cite{CD1,CD3}:
\begin{eqnarray}
U_{\vec{p},\sigma}(t,\vec{x}\,)&=& i N \left(\frac{p}{\omega}\right)^{\nu_-}(\omega t)^2\left(
\begin{array}{c}
e^{\pi \mu/2}H^{(1)}_{\nu_{-}}(-p t) \,
\xi_{\sigma}\\
e^{-\pi \mu/2}H^{(1)}_{\nu_{+}}(-p t) \,\frac{\vec{\sigma}\cdot\vec{p}}{p}\,\xi_{\sigma}
\end{array}\right)
e^{i\vec{p}\cdot\vec{x}}\label{Ups}\\
V_{\vec{p},\sigma}(t,\vec{x}\,)&=&-i N \left(\frac{p}{\omega}\right)^{\nu_+}(\omega t)^2 \left(
\begin{array}{c}
e^{-\pi \mu/2}H^{(2)}_{\nu_{-}}(-p t)\,\frac{\vec{\sigma}\cdot\vec{p}}{p}\,
\eta_{\sigma}\\
e^{\pi \mu/2}H^{(2)}_{\nu_{+}}(-p t) \,\eta_{\sigma}
\end{array}\right)
e^{-i\vec{p}\cdot\vec{x}}\,.\label{Vps}
\end{eqnarray}
The notation $\sigma_i$ stands for the Pauli matrices, while $H_{\nu_{\pm}}^{(1,2)}$  are the Hankel functions of
indices $\nu_{\pm}=\frac{1}{2}\pm i\mu$ with $\mu=\frac{m}{\omega}$. The
normalization constant
\begin{equation}
N=\frac{1}{(2\pi)^{3/2}}\frac{\sqrt{\pi}}{2}\,,
\end{equation}
assures the ortonormalization relations \cite{CD1}
\begin{eqnarray}
&&\left<U_{\vec{p},\sigma},U_{\vec{p}^{\,\prime},\sigma^{\prime}}\right>_D=
\left<V_{\vec{p},\sigma},V_{\vec{p}^{\,\prime},\sigma^{\prime}}\right>_D=
\delta_{\sigma\sigma^{\prime}}\delta^3 (\vec{p}-\vec{p}^{\,\prime})\,,
\label{orto1}\\
&&\left<U_{\vec{p},\sigma},V_{\vec{p}^{\,\prime},\sigma^{\prime}}\right>_D=
\left<V_{\vec{p},\sigma},U_{\vec{p}^{\,\prime},\sigma^{\prime}}\right>_D=
0\,,\label{orto2}
\end{eqnarray}
that give the inversion formulas,
\begin{equation}\label{invD}
a(\vec{p},\sigma)=\left<U_{\vec{p},\sigma},\psi\right>_D\,, \quad
a^c(\vec{p},\sigma)=\left<\psi,V_{\vec{p},\sigma}\right>_D\,.
\end{equation}

The above properties do not depend on the concrete choice of the Pauli spinors $\xi_{\sigma}$ and $\eta_{\sigma}= i\sigma_2 (\xi_{\sigma})^{*}$ as long as these are correctly normalized obeying $\xi^+_{\sigma}\xi_{\sigma'}=\eta^+_{\sigma}\eta_{\sigma'}=\delta_{\sigma\sigma'}$.
In general, we can project the spin on an arbitrary direction, which can be either dependent or independent on $\vec{p}$. The simplest cases are of the { momentum-spin} basis  and momentum-helicity basis \cite{CD1} (presented in Appendix A).  The fundamental spinors of the momentum-helicity basis can be written by replacing  $\xi_{\sigma}\to \xi_{\sigma}(\vec{p})$ and  $\eta_{\sigma}\to \eta_{\sigma}(\vec{p})$ in Eqs. (\ref{Ups}) and (\ref{Vps}) and using (\ref{heli}).

We must specify that the fundamental spinors (\ref{Ups}) and (\ref{Vps}) have  new phase factors different from those of Ref. \cite{CD1}. These are introduced in order to determine the limits
\begin{eqnarray}
\lim_{p \to 0} U_{\vec{p},\sigma}(t,\vec{x}\,)&=& U_{0,\sigma}(t)=  N_0 (-\omega t)^{\frac{i}{\omega} E_0^+
}\left(
\begin{array}{c}
\xi_{\sigma}\\
 0
\end{array}\right)\,,\label{cucu}\\
\lim_{p \to 0} V_{\vec{p},\sigma}(t,\vec{x}\,)&=& V_{0,\sigma}(t)= N_0^* (-\omega t)^{\frac{i}{\omega} E_0^-}
\left(
\begin{array}{c}
0\\
\eta_{\sigma}
\end{array}\right)\label{mucu}\,,
\end{eqnarray}
that hold in the momentum-spin basis where we can use (\ref{limits}). We find thus the fundamental spinors of the {\em natural rest frame} \cite{CD3} that represent the energy eigenspinors corresponding to the particle and respectively antiparticle {\em rest energies}, $E_0^{\pm}=\pm\,m-\frac{3i\omega}{2}$, whose last term is due to the decay produced by the de Sitter expansion \cite{CCC}. The normalization constant
\begin{equation}
N_0=\frac{e^{\frac{\pi\mu}{2}-i\mu\ln
2}}{(2\pi)^2}\,\Gamma\left(\textstyle{\frac{1}{2}}-i\mu\right)\,,
\end{equation}
is also interesting since its modulus $|N_0|=
(2\pi)^{-\frac{3}{2}}\left(1+e^{-2\pi\mu}\right)^{-\frac{1}{2}}$ could deal with some thermodynamic interpretations. The polarizations $\sigma=\pm\frac{1}{2}$ represent the spin projections on the third axis of the non-holonomic {\em rest} frame, which in our gauge (\ref{tt}) is parallel to the natural rest frame.  We can conclude that the mode separation is performed here just as in the Minkowskian QED defining thus an {\em adiabatic} vacuum state.  Notice that in the momentum-helicity basis, the state with $\vec{p}=0$ does not make sense and, therefore, in this representation we can not speak about rest frames.

The vacuum stability allows us to perform the {\em canonical} quantization assuming that the electron $(a,a^{\dagger})$ and positron $(a^c,{a^c}^{\dagger})$ field operators satisfy the non-vanishing anti-commutators \cite{CD1}
\begin{equation}\label{acom} \{a(\vec{p},\sigma),
a^{\dagger}({\vec{p}}^{\,\prime},\sigma^{\prime})\}= \{{a^c}(\vec{p},\sigma),
{a^c}^{\dagger}({\vec{p}}^{\,\prime},\sigma^{\prime})\}=
\delta_{\sigma\sigma^{\prime}}\delta^3 (\vec{p}-{\vec{p}}^{\,\prime})\,,
\end{equation}
so that the equal-time anti-commutator takes the canonical form \cite{CD1}
\begin{equation}\label{Detime}
\{ \psi(t,\vec{x}),\, \overline{\psi}(t, \vec{x}^{\,\prime})\}=
(-\omega t)^3\gamma^0 \delta^{3}(\vec{x}-\vec{x}^{\,\prime})\,.
\end{equation}
In general, for $t\not= t'$, the partial anti-commutator functions,
\begin{equation}
S^{(\pm)}(x,x')=
i\{ \psi^{(\pm)}(x),\,
\overline{\psi}^{(\pm)}(x')\}\,,
\end{equation}
and the total one
\begin{eqnarray}\label{defS}
S(x,x')&=& S^{(+)}(x,x')+ S^{(-)}(x,x')\nonumber\\
&=&i \int d^3 p \sum_{\sigma}\left[
U_{\vec{p},\sigma}(x)U^{+}_{\vec{p},\sigma}(x')+
V_{\vec{p},\sigma}(x)V^{+}_{\vec{p},\sigma}(x')
\right]\,,\label{Sfunc}
\end{eqnarray}
are solutions of the Dirac equation in both their sets of coordinates. They  help us to define the retarded ($R$) and advanced ($A$) Green functions as well as the Feynman propagator in standard manner as \cite{CD1},
\begin{eqnarray}
\tilde S^{R}(x,x')&=&\theta(t-t')S(x,x')\,,\\
\tilde S^{A}(x,x')&=&-\theta(t'-t)S(x,x')\,,\\
\tilde S^F(x,x')&=&
i\left<0\right|T[\psi(x)\overline{\psi}(x')]\left|0\right>
=\theta(t-t')S^{(+)}(x,x')\nonumber\\
&&-\theta(t'-t)S^{(-)}(x,x')\,.
\end{eqnarray}
All these Green functions obey
\begin{equation}
E_{D}(x)\tilde S^{F/R/A}(x,x')=
-\frac{1}{\sqrt{g(x)}}\, \delta^{4}(x-x')\,.
\end{equation}

The Noether theorem provides us with the  principal conserved observables of the quantum theory which are the electric charge,
\begin{equation}\label{Dcharge}
{\cal Q}[\psi]=:\langle \psi,\psi\rangle:=\int d^3x\,(-\omega t)^{-3} :\overline{\psi}\gamma^0 \psi :\,,
\end{equation}
and the $so(1,4)$ generators  associated to the Killing vectors of $M_{\omega}$. We have shown that any generator $X$ of the spinor
CR gives rise to the one-particle operator \cite{CD1}
\begin{equation}\label{opo}
{\cal X}[\psi]=:\left<\psi, (X\psi)\right>_D:
\end{equation}
calculated respecting the normal ordering  of the operator products (::) \cite{SW1}. All these operators form a representation the $so(1,4)$ algebra, commute with ${\cal Q}[\psi]$, and have the properties,
\begin{equation}\label{algXX}
\left[{\cal X}[\psi], \psi(x)\right]=-(X\psi)(x)\,, \quad
\left[{\cal X}[\psi], {\cal Y}[\psi]\right]=:\left<\psi, ([X,Y]\psi)\right>_D: \,.
\end{equation}
In particular,  the momentum and energy operators read
\begin{eqnarray}
\vec{\cal P}[\psi]&=&\int d^3 x  (-\omega t)^{-3}:\overline{\psi}\gamma^0 \left(-i\vec{\partial}\psi\right):\,,\label{PDir}\\
{\cal H}[\psi]&=&\int d^3x  (-\omega t)^{-3} : \overline{\psi}\gamma^0 {H}_D\psi :\,.\label{HDir}
\end{eqnarray}
The mode expansions of these operators were studied  in both the momentum and energy representations \cite{CD1,CD3}.

\subsection{The free Maxwell field}

The theory of the free Maxwell field minimally coupled to the de Sitter gravity  is governed by the action
\begin{equation}\label{action}
{\cal S}[A]=\int d^{4}x \sqrt{g}\,{\cal L}_{M}(A)=-\frac{1}{4}\int d^{4}x
\sqrt{g}\,F_{\mu \nu  }F^{\mu \nu  } ,
\end{equation}
where $A$ is the (electromagnetic) potential and $F_{\mu \nu  }=\partial_{\mu } A_{\nu}-\partial_{\nu } A_{\mu }$ is the field strength.
Now the canonical variables are the covariant components $A_{\mu}$ carrying natural indices. These components transform under isometries usually as any vector while the local components $\hat A_{\hat\alpha}=e_{\hat\alpha}^{\mu}A_{\mu}$ are those transforming according to the CR induced by the vector representation of the $SL(2,\Bbb C)$ group \cite{EPL,CCC}. For this reason the action of the energy operator upon the covariant components  \cite{Max},
\begin{equation}\label{hamiM}
(H A)_{\mu}= -i\omega (t\partial_t+x^i\partial_i+1)A_{\mu}\,.
\end{equation}
has a supplemental term, which does not appear in Eq. (\ref{hami}).

The free field equation derived from this action is conformally invariant but the Lorentz condition keeps this property only in the Coulomb gauge \cite{Max}. Therefore, in order to preserve the conformal invariance of the whole Maxwell theory we fix this gauge in which $A_0=0$ and $\partial_i A_i=0$. Then, the free Maxwell equation on $M_{\omega}$,
\begin{equation}\label{EMax}
E_M(x)A_i(x)=\frac{1}{\sqrt{g(x)}}\,(\partial_t^2-\Delta)A_i(x)=0\,,
\end{equation}
can be solved in momentum-helicity basis \cite{Max} where the potential
has the expansion
\begin{equation}\label{Max}
\vec{A}(x)=\int d^3k
\sum_{\lambda}\left[\vec{w}_{\vec{k},\lambda}(x) \alpha({\vec
k},\lambda)+\vec{w}_{\vec{k},\lambda}(x)^* \alpha^{\dagger}({\vec
k},\lambda)\right]\,,
\end{equation}
in terms of the modes functions \cite{Max},
\begin{equation}\label{fk}
\vec{w}_{\vec{k},\lambda}(t,\vec{x}\,)=
\frac{1}{(2\pi)^{3/2}}\frac{1}{\sqrt{2k}}\,e^{-ikt+i{\vec
k}\cdot {\vec x}}\,\vec{\varepsilon}_{\lambda} (\vec{k})\,,
\end{equation}
depending on the momentum $\vec{k}$ ($k=|\vec{k}|$) and helicity $\lambda=\pm 1$ of the polarization vectors ${\vec\varepsilon}_{\lambda}({\vec k})$ in Coulomb gauge (given in Appendix A). According to our asymptotic prescription these particle modes functions are progressive waves, as in the flat case, defining thus  the {\em conformal} vacuum \cite{BD}.

According to  Eq. (\ref{orto}) we can verify that the modes functions (\ref{fk}) are orthonormal,
\begin{eqnarray}
&&\left<w_{\vec{k},\lambda},w_{\vec{k}\,'\lambda'}\right>_M
=-\left<w^*_{\vec{k},\lambda},w^*_{\vec{k}\,',\lambda'}\right>_M=\delta_{\lambda\lambda'}\delta^3(\vec{k}-\vec{k}\,')\,,\\
&&\left<w_{\vec{k},\lambda},w^*_{\vec{k}\,',\lambda'}\right>_M=
\left<w^*_{\vec{k},\lambda},w_{\vec{k}\,',\lambda'}\right>_M=0\,,
\end{eqnarray}
with respect to the relativistic scalar product of the Maxwell theory
\begin{equation}
\left<w,w'\right>_M=i \int d^3 x\, w^*(t,\vec{x})_i\stackrel{\leftrightarrow}{\partial_t}w'(t,\vec{x})_i
\end{equation}
where we denote $f\stackrel{\leftrightarrow}{\partial} g=f\partial g-g\partial f$. Consequently, we find the simple inversion formula
\begin{equation}\label{invM}
\alpha(\vec{k},\lambda)=\left<w_{\vec{k},\lambda},A\right>_M\,.
\end{equation}

The conformal invariance of the Maxwell theory on $M_{\omega}$  enables us to perform the second quantization in Coulomb gauge as in special relativity  assuming that the photon field operators satisfy  \cite{Max}
\begin{equation}\label{com1}
[\alpha({\vec k},\lambda), \alpha^{\dagger}({\vec k}^{\,\prime},\lambda ')]
=\delta_{\lambda\lambda '} \delta^3 ({\vec k}-{\vec k}^{\,\prime})\,.
\end{equation}
Then the  Hermitian field $A=A^{\dagger}$ and its momentum density $\pi^j=\partial_{t}A_j$ satisfy the {\em canonical} rule
\begin{equation}\label{Aetime}
[ A_i(t,\vec{x}),\pi^j(t,\vec{x}\,')]=[ A_i(t,\vec{
x}),\partial_{t}A_j(t,\vec{x}\,')]=i\,\delta^{tr}_{ij}(\vec{x}-\vec{x}\,')\,,
\end{equation}
where   $\delta^{tr}_{ij}(\vec{x})=(\delta_{ij}-\partial_k\partial_i\Delta^{-1}_x) \delta^3(\vec{x})$ is the transverse $\delta$-function \cite{BDR,Max}.

As in the flat case, the Green functions are related to the partial commutator functions of positive or negative frequencies,
\begin{equation}
D_{ij}^{(\pm)}(x-x')= i[A_i^{(\pm)}(x),A_j^{(\pm)\,\dagger}(x')]
\end{equation}
and the total one, $D_{ij}=D^{(+)}_{ij}+D^{(-)}_{ij}$. These functions have simple mode expansions
\begin{equation}\label{defD}
D_{ij}^{(+)}(x-x')=\left[D_{ij}^{(-)}(x-x')\right]^*=i \int d^3 k \sum_{\lambda} w_{\vec{k},\lambda}(x)_i\,w^*_{\vec{k},\lambda}(x')_j
\end{equation}
showing that these satisfy the field equation having vanishing divergences in both the sets of variables.  With their help we can construct different {\em transverse} Green functions, $\tilde D_{ij}(x)=\tilde D_{ji}(x)$, which obey
\begin{equation}\label{KGG}
E_M(x)\tilde D_{ij}(x-x')=\frac{1}{\sqrt{g(x)}}\,\delta(t-t')\delta^{tr}_{ij}(\vec{x}-\vec{x}\,')
\end{equation}
and $\partial_{i}\tilde D_{\cdot j}^{i \cdot}(x)=0$. Of a special interest
are the retarded, $\tilde D^R_{ij}(x)=\theta(t)D_{ij}(x)$, and advanced,
$\tilde D^A_{ij}(x)=-\theta(-t)D_{ij}(x)$, transverse Green functions.
The transverse Feynman propagator,
\begin{eqnarray}
&&\tilde D^F_{ij}(x-x')= i\langle 0|T[A_i(x) A_j(x')]\,|0\rangle\nonumber\\
&&= \theta (t-t')
D_{ij}^{(+)}(x-x')-\theta(t'-t)D_{ij}^{(-)}(x- x')\,,
\end{eqnarray}
is defined as a  causal Green function. It is not difficult to verify that all these functions satisfy Eq. (\ref{KGG}). Thus we conclude  that the Green functions of the free Maxwell  field in the chart $\{t,\vec{x}\}$  have the same forms and properties as those of the Maxwell theory in Minkowski spacetime.

However, physically speaking, all these apparent similarities resulted from the conformal invariance are merely formal because of the special definition of the conformal time. This can be observed by applying the Noether theorem which turns out the conserved one-particle operators \cite{Max}
\begin{equation}\label{mopo}
{\cal X}[A]=\frac{1}{2}\,:\langle A, (X A)\rangle_M:
\end{equation}
produced by the isometry generators, $X$.  The obvious
algebraic properties
\begin{equation}\label{malgXX}
\left[{\cal X}[A], A_i(x))\right]=-(X A)_i(x)\,, \quad \left[{\cal X}[A], {\cal
Y}[A]\,\right]=\frac{1}{2}\,:\langle A, ([X,Y]A)\,\rangle_M:
\end{equation}
are due to the canonical quantization adopted here. Hereby we can write the  momentum and energy operators,
\begin{eqnarray}
\vec{\cal P}[ A]&=&\int d^3 x :\vec{E}\land\vec{B}:\,,\label{PMax}\\
{\cal H}[A]&=&\int d^3x :\left[(-\omega t) \frac{1}{2}\,\left(\vec{E}\,^2+\vec{B}\,^2\right)+\omega \vec{x}\cdot \vec{E}\land\vec{B}-\frac{\omega}{2}\,\partial_t(\vec{A}\,^2)\right]:\,,\label{HMax}
\end{eqnarray}
in terms of electric, $\vec{E}$, and magnetic, $\vec{B}$, components of the field strength (denoted as in Ref. \cite{BDR}). Thus we see that now the energy operator is dramatically different from that of the flat case, even though the momentum one remains the same. We specify that the last term  of Eq. (\ref{HMax}) correspond to the supplemental term of Eq. (\ref{hamiM}). The properties of these operators were analysed using expansions in momentum or energy representations \cite{Max}.

\section{The de Sitter QED}

Let us consider now the covariant fields $\Psi$ and ${\cal A}$, minimally coupled to the gravity of $M_{\omega}$, interacting between themselves according to the QED action
\begin{equation}
{\cal S}_{QED}=\int d^4 x\sqrt{g}\, :\left[ {\cal L}_D(\Psi)+{\cal L}_{M}({\cal A})+{\cal L}_{int}(\Psi,{\cal A})\right]:
\end{equation}
whose interaction term,
\begin{equation}\label{Lint}
{\cal L}_{int}(\Psi,{\cal A})=-e_0
\overline{\Psi}(x)\gamma^{\hat\mu}e^{\nu}_{\hat\mu}(x){\cal A}_{\nu}(x)\Psi(x)\,,
\end{equation}
defines the minimal electromagnetic coupling given by the electric charge $e_0$. This Lagrangian density is tetrad-gauge invariant and, in addition, remains invariant under the $U(1)_{em}$  transformations of the electromagnetic gauge. Therfore, the action which is invariant  under the $SO(1,4)$ isometries remains unaffected by these gauge transformations.  In general, the isometries may change the electromagnetic gauge so that this has to be corrected after each isometry.  This procedure is similar to that of the tetrad-gauge transformations associated to isometries  that preserve the gauge fixing of the CRs  \cite{ES}. Obviously, in this framework the quantization must be performed only after fixing both the tetrad-gauge and the  elecrtomagnetic one.

The interacting fields satisfy the system of coupled equations
\begin{eqnarray}
E_D(x)\Psi(x)&=&e_0\, \gamma^{\hat\mu}e^{\nu}_{\hat\mu}(x):{\cal A}_{\nu}(x)\Psi(x):\,,\label{EQ1}\\
\partial_{\nu}\left[\sqrt{g}g^{\nu\alpha}g^{\mu\beta}(\partial_{\alpha}{\cal A}_{\beta}-\partial_{\beta}{\cal A}_{\alpha})\right](x)&=&-\sqrt{g(x)}\,e^{\mu}_{\hat\nu}(x)\, {\cal J}^{\hat\nu}(x)\,,  \label{EQ2}
\end{eqnarray}
where
\begin{equation}\label{JJJ}
{\cal J}^{\hat\nu}(x)=-e_0 :\overline{\Psi}(x)\gamma^{\hat\nu}\Psi(x):
\end{equation}
are the components of the conserved current density in the local frame $\{e\}$ where these obey $\partial_{\mu}(\sqrt{g}\,e^{\mu}_{\hat\alpha}{\cal J}^{\hat\alpha})=0$. The solution of these equations are quantum fields whose properties are strongly dependent on the electromagnetic gauge we choose.

\subsection{Quantization in Coulomb gauge}

The canonical quantization of the free Maxwell field was successfully performed in Coulomb gauge where the Lorentz condition becomes conformally invariant. This means that it is useful to maintain this gauge for the interacting fields assuming that  $\partial_i {\cal A}_i=0$.  Notice that this is always possible setting ${\cal A}_{\mu}\to {\cal A}_{\mu}-\partial_{\mu}\Delta^{-1}\partial_i{\cal A}_i$.
In this gauge Eqs. (\ref{EQ2}) carry out the static equation
$\Delta{\cal A}_0(x)=(-\omega t)^{-3} {\cal J}^0(x)$  giving the Coulomb potential
\begin{equation}\label{Coul}
{\cal A}_0(t,\vec{x})=\frac{1}{(-\omega t)^3}\frac{e_0}{4\pi}\int \frac{d^3 x'}{|\vec{x}\,'-\vec{x}|}\,:\overline{\Psi}(t,\vec{x}\,')\gamma^0\Psi(t,\vec{x}\,'):\,,
\end{equation}
and the transverse equation $E_M(x){\cal A}_k(x)=\omega t {\cal J}^k_{tr}(x)$ depending only on the transverse current density ${\cal J}^k_{tr}(x)={\cal J}^k(x)-\partial_k\partial_i\Delta^{-1}_x {\cal J}^i(x)$.

The conserved one-particle operators of the interacting fields can be derived applying the Noether theorem piece by piece instead of using simple formulas as (\ref{opo}) and (\ref{mopo}). Here we restrict ourselves to consider only the electric charge ${\cal Q}[\Psi]$ given by Eq. (\ref{Dcharge}) and the momentum and energy operators  that read now
\begin{eqnarray}
\vec{\cal P}[\psi;{\cal A}]&=&\int d^3 x :\left[ (-\omega t)^{-3}\overline{\Psi}\gamma^0 (-i\vec{\partial}\Psi)+ \vec{\cal E}\land\vec{\cal B}\right]:\,,\label{PENE}\\
{\cal H}[\psi;{\cal A}]&=&\int d^3x :\left[ (-\omega t)^{-3} \overline{\Psi}\gamma^0 {H}_D\Psi - (\omega t)^{-2}\left({\cal J}^0{\cal A}_0+\vec{\cal J}\cdot \vec{\cal A}\right) \right]:\nonumber\\
&+&\int d^3 x :\left\{(-\omega t) \left[\frac{1}{2}\,\left(\vec{\cal E}\,^2+\vec{\cal B}\,^2\right)+{\cal A}_0 \Delta{\cal A}_0 \right]\right.\nonumber\\
&& \hspace*{46mm}\left.+\,\omega\, \vec{x}\cdot \vec{\cal E}\land\vec{\cal B}-\frac{\omega}{2}\,\partial_t(\vec{\cal A}\,^2)\right\}:\,,\label{HENE}
\end{eqnarray}
where $\vec{\cal E}$ and $\vec{\cal B}$ are the electric and magnetic fields generated by the potential ${\cal A}$. The static equation allows us to drop out the component ${\cal J}^0$ and write  ${\cal H}[\psi;{\cal A}]={\cal H}[\Psi]+{\cal H}[{\cal A}]+{\cal H}_{int}[\psi;{\cal A}]$ separating thus the interaction Hamiltonian
\begin{equation}\label{HintPA}
{\cal H}_{int}[\psi;{\cal A}](t)=e_0 (\omega t)^{-2}\int d^3 x  :\overline{\Psi}(t,\vec{x})\,\vec{\gamma}\cdot \vec{\cal A}(t,\vec{x})\Psi(t,\vec{x}):\,,
\end{equation}
from the kinetic part  given by Eqs. (\ref{HDir}) and (\ref{HMax}). We demonstrate thus that the interaction Hamiltonian in Coulomb gauge  does not depend on ${\cal A}_0$, just as it happens in the Minkowskian QED. Moreover, we observe that the momentum operator has no longer interaction terms since $\vec{\cal P}[\psi;{\cal A}]=\vec{\cal P}[\Psi]+\vec{\cal P}[{\cal A}]$ as it results from Eqs. (\ref{PDir}) and (\ref{PMax}). This fact suggests that there exists an evolution operator depending only on time and, consequently,  commuting with $\vec{\cal P}[\psi;{\cal A}]$  \cite{BDR}.

The above defined operators have correct actions only if we assume that the equal-time commutation and anti-commutation relations of the interacting fields are just those of the free fields, (\ref{Aetime}) and (\ref{Detime}). Therefore, we consider the non-vanishing relations
\begin{eqnarray}
\left\{ \Psi(t,\vec{x}),\, \overline{\Psi}(t, \vec{x}\,')\right\}&=&
(-\omega t)^3\gamma^0 \delta^{3}(\vec{x}-\vec{x}\,')\,,\\
\left[ {\cal A}_i(t,\vec{x}),\partial_{t}{\cal A}_j(t,\vec{x}\,')\right]&=&i\,\delta^{tr}_{ij}(\vec{x}-\vec{x}\,')\,,
\end{eqnarray}
supposing, in addition, that $\Psi$ commute at equal time with both, ${\cal A}$ and  $\partial_t{\cal A}$. Hereby we obtain the commutation relations of the Coulomb potential (\ref{Coul}) that read,
\begin{eqnarray}
\left[ {\cal A}_0(t,\vec{x}),{\cal A}_j(t,\vec{x}\,')\right]&=&\left[ {\cal A}_0(t,\vec{x}),\partial_{t}{\cal A}_j(t,\vec{x}\,')\right]=0\,,\\
\left[ {\cal A}_0(t,\vec{x}), \Psi(t,\vec{x}\,')\right]&=&-\frac{1}{(-\omega t)^3}\frac{e_0}{4\pi} \frac{1}{|\vec{x}\,'-\vec{x}|}\,\Psi(t,\vec{x}\,')\,.
\end{eqnarray}
These have the familiar form \cite{BDR} apart from the factor $(-\omega t)^{-3}$ that tends to 1 in the flat limit.

Now we can use the above relations and the field equations in Coulomb gauge in order to verify the desired relations
\begin{eqnarray}
[{\cal Q}[\Psi],\Psi]&=&-\Psi\,,\\
\left[\vec{\cal P}[\psi;{\cal A}],\Psi\right]&=&-\vec{P}\Psi=i\vec{\partial}\Psi\,\\
\left[{\cal H}[\psi;{\cal A}],\Psi\right]&=&-H\Psi=i\omega(t\partial_t+x^i\partial_i)\Psi \label{HamPsi}
\end{eqnarray}
and similarly, $[{\cal Q}[\Psi],{\cal A}]=0$ and
\begin{eqnarray}
\left[\vec{\cal P}[\psi;{\cal A}],{\cal A}\right]&=&-\vec{P}{\cal A}=i\vec{\partial}{\cal A}\,,\\
\left[{\cal H}[\psi;{\cal A}],{\cal A}\right]&=&-H{\cal A}=i\omega(t\partial_t+x^i\partial_i+1){\cal A}\,.
\end{eqnarray}
All the basis-generators of the $so(1,4)$ algebra can be written in terms of interacting fields having similar properties. They satisfy appropriate   algebraic rules and commute with ${\cal Q}[\Psi]$. Then it is not surprising to find that in the flat limit  we recover the Poincar\' e generators of the usual QED in Coulomb gauge \cite{CCC}.

\subsection{$in-out$ transition amplitudes}

The $in$ and $out$ fields are special free fields with  convenient asymptotic behaviours that  can be defined using appropriate Green functions. The starting point is the formal representation of the  solutions of the system (\ref{EQ1}), (\ref{EQ2}) that can be written in terms of free fields ($\psi$ and $A$) and Green functions ($\tilde S$ and $\tilde D_{ij}$) as
\begin{eqnarray}
\Psi(x)&=&\psi(x)-e_0\int d^4x'\sqrt{g(x')}\, \tilde S(x,x')\gamma^{\hat\mu}e^{\nu}_{\hat\mu}(x'):{\cal A}_{\nu}(x')\Psi(x'):\\
{\cal A}_k(x)&=&A_k(x)-\int d^4x'\sqrt{g(x')}\, \tilde D_{ki}(x-x'\,)e^i_{\hat\nu}(x'){\cal J}^{\hat\nu}(x')\,.\label{AD}
\end{eqnarray}
Notice that the fields $A$ and ${\cal A}$ remain simultaneously in Coulomb gauge since  the Green function $\tilde D_{ij}$ selects the transverse part of the current density in Eq. (\ref{AD}).

The above representations enable us to  define,
\begin{eqnarray}
\sqrt{z_2}\psi^{in/out}(x)&=&\Psi(x)+\int d^4 x' \sqrt{g(x')}\tilde S^{R/A}(x,x')E_D(x')\Psi(x')\,,\\
\sqrt{z_3}{A}_i^{in/out}(x)&=&{\cal A}_i(x)-\int d^4 x' \sqrt{g(x')}\tilde D^{R/A}_{ij}(x-x')E_M(x') {\cal A}_j(x')\,,
\end{eqnarray}
as in the flat case. These fields satisfy the free field equations (\ref{EDir}) and (\ref{EMax}) and desired asymptotic conditions,
\begin{eqnarray}
\lim_{t\to -\infty}\left[\sqrt{z_2}\psi^{in}(x)-\Psi(x)\right]=0\,,&&
\lim_{t\to 0}\left[\sqrt{z_2}\psi^{out}(x)-\Psi(x)\right]=0\,,\\
\lim_{t\to -\infty}\left[\sqrt{z_3}A_i^{in}(x)-{\cal A}_i(x)\right]=0\,,&&
\lim_{t\to 0}\left[\sqrt{z_3}A_i^{out}(x)-{\cal A}_i(x)\right]=0\,,
\end{eqnarray}
which in the moving chart of proper time lead to the usual limits for $\hat t\to -\infty$ and $\hat t\to \infty$ respectively. The constants $z_2$ and $z_3$ are introduced for preserving the standard normalization of the free fields as defined in the previous section. These will be important pieces of the renormalization procedure \cite{BDR}.

Under such circumstances, it is obvious that the $in$ and $out$ free fields can be quantized in canonical manner so that their field operators in momentum representation, $a_{in}(\vec{p},\sigma)=\langle U_{\vec{p},\sigma},\psi^{in}\rangle_{D},...$ etc., obey the standard commutation or anti-commutation rules given by Eqs. (\ref{com1}) and (\ref{acom}). These operators generate two different bases of the Fock space whose state vectors are denoted by $|in;...\rangle$ and respectively $|out;...\rangle$. First of all we assume that there exists a unique vacuum state $|in;0\rangle=|out;0\rangle=|0\,\rangle$ that satisfies $a_{in}(\vec{p},\sigma)|0\,\rangle=...=a_{out}(\vec{p},\sigma)|0\rangle=...=0$. The creation operators give rise to $in$ or $out$ state vectors with different numbers of particles. For example, the $in$ vector of a state having $n_1$ electrons, $n_2$ positrons and $n_3$ photons reads
\begin{equation}
|in; \,n_1(\vec{p}\,\sigma)_-, n_2(\vec{p}\,'\,\sigma')_+,n_3 (\vec{k}\,\lambda)\rangle= {a_{in}}^{\dagger}(\vec{p},\sigma)^{n_1}
{a^c_{in}}^{\dagger}(\vec{p}\,',\sigma')^{n_2} \alpha_{in}^{\dagger}(\vec{k},\lambda)^{n_3}|0\rangle\,.
\end{equation}
The vacuum together with all the $in$ vectors constitute the $in$ momentum basis while the $out$ momentum basis can be build in a similar way. In general, these bases are different such that the transition coefficients $\langle out; \beta....|in; \alpha...\rangle$ are not trivial representing transition amplitudes between $in$ states at $t\to-\infty$ (or $\hat t\to -\infty$) and  $out$ states at $t=0$ (or $\hat t \to \infty$). Unfortunately, as in the flat case, these amplitudes can  be calculated only by using perturbations that require renormalization. For this reason it is crucial to adopt the hypothesis of stability of the one particle sector,
\begin{eqnarray}
\langle out;(\vec{p}\,\sigma)_{\pm}|in;(\vec{p}\,'\sigma')_{\pm}\rangle&=&\delta_{\sigma\sigma'}
\delta^3(\vec{p}-\vec{p}\,')\,,\\
\langle out;(\vec{k}\,\lambda)|in;(\vec{k}\,'\lambda')\rangle&=&\delta_{\lambda\lambda'}
\delta^3(\vec{k}-\vec{k}\,')\,.
\end{eqnarray}
since this represents the principal criterion for determining the (re)normaliza\-tion constants $z_2$ and $z_3$ \cite{BDR}.

The transition amplitudes can be evaluated by using the reduction formalism which is based on the fact that the $in$ and $out$ fields can be related among themselves as
\begin{eqnarray}
\sqrt{z_2}[\psi^{in}(x)-\psi^{out}(x)]&=&\int d^4 x' \sqrt{g(x')} S(x,x')E_D(x')\Psi(x')\,,\\
\sqrt{z_3}[A_i^{in}(x)-A_i^{out}(x)]&=&-\int d^4 x' \sqrt{g(x')} D_{ij}(x-x')E_M(x') {\cal A}_j(x')\,.
\end{eqnarray}
The functions $S=\tilde S^R-\tilde S^A$ and $D=\tilde D^R-\tilde D^A$ are just those defined by Eqs. (\ref{defS}) and (\ref{defD}) respectively. Therefore, we can use the inversion formulas (\ref{invD}) and (\ref{invM}) for obtaining the differences
\begin{eqnarray}
\sqrt{z_2}\left[a_{in}(\vec{p},\sigma)-a_{out}(\vec{p},\sigma)\right]&=&
i\int d^4 x \sqrt{g(x)}\, \overline{U}_{\vec{p},\sigma}(x)E_D(x)\Psi(x)\,,\\
\sqrt{z_2}\left[a^c_{in}(\vec{p},\sigma)-a^c_{out}(\vec{p},\sigma)\right]&=&
-i\int d^4 x \sqrt{g(x)}\, \overline{E_D(x)\Psi(x)}V_{\vec{p},\sigma}(x)\,,\\
\sqrt{z_3}\left[\alpha_{in}(\vec{k},\lambda)-\alpha_{out}(\vec{k},\lambda)\right]&=&
i\int d^4 x \sqrt{g(x)}\, w^*_{\vec{k},\lambda}(x)_i E_M(x){\cal A}_i(x)\,,\end{eqnarray}
that represent the starting point of the reduction formalism  which works here as in the flat case \cite{BDR}. For example, the reduction of an electron from the $out$ states gives
\begin{eqnarray}
&&\langle out;(\vec{p}\,\sigma)_-,\alpha|in;\beta\rangle\nonumber\\
&&\hspace*{12mm}=-\frac{i}{\sqrt{z_2}}
\int d^4 x \sqrt{g(x)}\, \overline{U}_{\vec{p},\sigma}(x)E_D(x)\langle out;\alpha|\Psi(x)|in;\beta\rangle\,.
\end{eqnarray}
The complete set of reduction rules of the de Sitter QED is presented in Ref. \cite{CR}. Using this mechanism we can reduce all the particles from the $out$ and $in$ states arriving to a multiple integral involving a generalized Green function  defined as the vacuum expectation value of a chronological product of interacting fields. These Green functions, called often $\tau$-functions  \cite{BDR}, can be calculated using perturbations.

\subsection{Perturbations}

The basic assumption of the perturbation theory is that the interacting fields, $\Psi$ and ${\cal A}$, can be related to a system of free fields, $\psi$ and $A$, as
\begin{eqnarray}
\Psi(t,\vec{x})&=&U^{\dagger}(t)\psi(t,\vec{x})U(t)\,,\label{UU1}\\
{\cal A}_i(t,\vec{x})&=&U^{\dagger}(t)A_i(t,\vec{x})U(t)\,,\label{UU2}
\end{eqnarray}
with the help of the unitary operator $U(t)$ depending only on time as  the translation invariance suggests. Furthermore, by using the well-known method and arguments of the perturbation theory \cite{BDR} we  apply the operator (\ref{hami}) on Eq. (\ref{UU1}) obtaining,  according to Eqs. (\ref{algXX}a) and (\ref{HamPsi}),  the identity
\begin{equation}
\left[{\cal H}[\psi]+{\cal H}[A],\psi\right]=i\omega t \left[(\partial_t U)U^{\dagger},\psi\right]+\left[{\cal H}[\psi,A],\psi\right]
\end{equation}
and a similar one for the Maxwell field. We observe that the operator  ${\cal H}[\psi,A]-{\cal H}[\psi] -{\cal H}[A]={\cal H}_{int}[\psi,A]$ is in fact the interaction Hamiltonian as given by Eq. (\ref{HintPA}) but with the free fields replacing the interacting ones. Thus we deduce that the evolution operator $U(t,t')=U(t)U^{\dagger}(t')$ satisfies the differential equation
\begin{equation}
\partial_t U(t,t')=\frac{i}{\omega t}{\cal H}_{int}[\psi,A](t)U(t,t')
\end{equation}
with the initial condition $U(t,t)=I$. This is equivalent to the integral representation
\begin{equation}
U(t,t')=I+i \int^t_{t'}\frac{d\tau}{\omega \tau}  {\cal H}_{int}[\psi,A](\tau)\,U(\tau,t')
\end{equation}
which generates the series of perturbations. Hereby it turns out that
the generalized Green functions can be expanded in terms of free fields as
\begin{equation}\label{GGR}
\langle 0| T\left(\Psi(x_1)\overline{\Psi}(x_2){\cal A}_i(x_3)...\right)|0\rangle =
\frac{\langle 0| T\left(\psi(x_1)\overline{\psi}(x_2)A_i(x_3)...{\bf S}\right)|0\rangle}{\langle 0|{\bf S}|0\rangle}
\end{equation}
where
\begin{equation}\label{Smatrix}
{\bf S}=U(0,-\infty)=T\, \exp \left[-ie_0\int \frac{d^4x}{(-\omega t)^3}\,
:\overline{\psi}(x)\,\vec{\gamma}\cdot\vec{A}(x)\,\psi(x): \right]
\end{equation}
is the series which can be used in applications. This result is important since it suggests the general formula,
\begin{equation}\label{Smatrix1}
{\bf S}=T\, \exp \left[-ie_0\int d^4x \sqrt{g(x)}\,
:\overline{\psi}(x)\,{\gamma}^{\hat\mu} e_{\hat\mu}^j(x) {A}_j(x)\,\psi(x): \right]\,,
\end{equation}
that holds  in both the moving charts, $\{t,\vec{x}\}$ and $\{\hat t,\vec{x}\}$, with {\em any} tetrad-gauge as long as we preserve the Coulomb gauge. Notice that in the flat limit ($g\to 1$, $e_0^j\to 0$ and $e_i^j\to \delta_i^j$) we recover the usual formula of the scattering operator in this gauge and flat space-time.

Finally, we remind the reader that the functions of the form (\ref{GGR}) can be expressed in terms of Feynman propagators considering all the possible t-contractions as the Wick theorem states. Then the numerators of these functions can be split in connected parts multiplied just by the vacuum expectation value $\langle0|{\bf S}|0\rangle$ such that after simplification we remain only with the connected terms giving the physical transition amplitudes \cite{BDR}.

\section{First order transition amplitudes}

We consider now the simplest examples of amplitudes that can be calculated in the de Sitter QED. As mentioned above, we meet here some effects which have non-vanishing amplitudes even in the first order of perturbations. These involve only three particles, a photon and two Dirac particles, which can appear in the $in$ or $out$ states. The allowed transitions are: one electron ($e^-$) or positron ($e^+$) emitting or absorbing one photon ($\gamma$), one photon pair
creation or annihilation ($\gamma\to e^-+e^+$ and $e^-+e^+\to \gamma$), the creation from the QED vacuum ($vac$) of the triplet $e^-+e^+ + \gamma$ and the annihilation of this triplet into the same vacuum.

The electromagnetic particle creation  $vac\to e^-+e^+ + \gamma$  was studied  in \cite{IITsar} where the total amplitude was calculated between an $in$ state at $t\to -\infty$ and the $out$ state at $t\to \infty$. Thus the conformal time $t$ covers the expansion period $(-\infty,0]$ followed by a symmetrical contraction, for $t\in [0,\infty)$. In this way the contraction cancels the effects due to the expansion, vanishing thus the total transition amplitude \cite{IITsar}. In our opinion, the contributions of the expansion and contraction periods must be treated separately considering the $out$ state at $t=0$ in the scenario of the expanding universe. Consequently, we find that the electromagnetic  particle creation has a non-vanishing amplitudes as we present below.

\subsection{Related amplitudes}

In the first order of perturbations we can take $z_2=z_3=1$. By using then the above presented formalism we find two types of simple amplitudes.

{\em I. The first type} is represented by the amplitude of the photon emission $e^-\to e^- + \gamma$ that reads
\begin{eqnarray}
&&\langle (\vec{p}\,'\sigma')_-, (\vec{k}\,\lambda)|{\bf S}_1|(\vec{p}\,\sigma)_-\rangle\nonumber\\
&&\hspace*{30mm}=ie_0\int\frac{d^4x}{(\omega
t)^3}\,\overline{U}_{{\vec{p}\,}',\sigma'}(x)\, \vec{\gamma}\cdot
\vec{w}_{\vec{k},\lambda}(x)^* U_{\vec{p},\sigma}(x)\,.
\end{eqnarray}
When the photon is emitted by a positron we have to replace
${U}_{{\vec{p}\,}',\sigma'}\to {V}_{\vec{p},\sigma}$ and
${U}_{\vec{p},\sigma}\to {V}_{{\vec{p}\,}',\sigma'}$. Moreover, if
we replace $\vec{w}^{\,*}\to \vec{w}$ we obtain the amplitudes of
the transitions $e^-+\gamma \to e^-$ and $e^++\gamma \to e^+$ in
which a photon is absorbed.

{\em II. The second type} appears in the cases of the pair creation,
$\gamma \to e^- + e^+$, and annihilation, $e^- + e^+ \to \gamma$, when we
find the related amplitudes
\begin{eqnarray}\label{pair}
&&\langle(\vec{p}\,\sigma)_-,(\vec{p}\,'\sigma')_+ |{\bf S}_1|(\vec{k}\,\lambda)
\rangle =-\langle (\vec{k}\,\lambda)|{\bf S}_1| (\vec{p}\,\sigma)_-,
(\vec{p}\,'\sigma')_+
 \rangle^*\nonumber\\
&&\hspace*{30mm}=ie_0\int\frac{d^4x}{(\omega
t)^3}\,\overline{U}_{\vec{p},\sigma}(x)\, \vec{\gamma}\cdot
\vec{w}_{\vec{k},\lambda}(x) V_{{\vec{p}\,}',\sigma'}(x)\,,
\end{eqnarray}
observing that the amplitudes of the transitions  $vac \to e^- + e^+ +\gamma$ and $e^- + e^+ +\gamma\to vac$  can be written
by replacing $\vec{w}\to \vec{w}^{\,*}$ in Eq. (\ref{pair}).

These amplitudes can be easily calculated in momentum-helicity basis according to Eqs. (\ref{Ups}), (\ref{Vps}), rewritten in this basis, and  Eq. (\ref{fk}). After a few manipulations we obtain
\begin{eqnarray}
&&\langle (\vec{p}\,'\sigma')_-,
(\vec{k}\,\lambda)|{\bf S}_1|(\vec{p}\,\sigma)_-\rangle \nonumber\\
&&\hspace*{18mm}=\delta^3(\vec{p}-{\vec{p}\,}'-\vec{k})\,\frac{ i e_0}{16\sqrt{\pi}}\sqrt{\frac{p
p'}{k}}\,\left(\frac{p}{p'}\right)^{i\mu}\xi_{\sigma'}^+({\vec{p}\,}\,')
\,\vec{\sigma}\cdot
{\vec{\varepsilon}_{\lambda}(\vec{k})}^*\xi_{\sigma}({\vec{p}}\,)\nonumber\\
&&\hspace*{22mm}\times\left[{\rm sign}(\sigma)\, I^{(2,1)}_+(p,p',-k)+{\rm
sign}(\sigma')\,I^{(2,1)}_-(p,p',-k)\right]\,, \label{ampl1}\\
&&\langle (\vec{p}\,\sigma)_-, (\vec{p}\,'\sigma')_+ |{\bf S}_1|(\vec{k}\,\lambda)\rangle \nonumber\\
&&\hspace*{18mm}=-\delta^3(\vec{p}+\vec{p}\,'-\vec{k})\,
\frac{i e_0}{16\sqrt{\pi}}\sqrt{\frac{p p'}{k}}\,
\left(\frac{p p'}{\omega^2}\right)^{i\mu}\,\xi_{\sigma}^+(\vec{p}\,)
\,\vec{\sigma}\cdot
\vec{\varepsilon}_{\lambda}(\vec{k})\,\eta_{\sigma'}({\vec{p}\,}')\nonumber\\
&&\hspace*{22mm}\times\left[e^{\pi\mu} I^{(2,2)}_+(p,p',k)-{\rm
sign}(\sigma\sigma')\,e^{-\pi\mu}I^{(2,2)}_-(p,p',k)\right]\label{ampl2}\,,
\end{eqnarray}
where we denote
\begin{equation}\label{int0}
I^{(a,b)}_\pm(p,p',q)=\int_{0}^{\infty}ds\, s\,
H^{(a)}_{\nu_\pm}(sp)H^{(b)}_{\nu_\pm}(sp')\, e^{i q s}\,,\quad a,b=1,2\,,
\end{equation}
the time integrals of Hankel functions in the new variable $s=-t$.
The obvious properties,
\begin{eqnarray}
I^{(a,b)}_\pm(p,p',q)&=&I^{(b,a)}_\pm(p',p,q)\,,\quad a,b=1,2\,,\\
{I^{(1,1)}_\pm(p,p',q)}^*&=&I^{(2,2)}_\mp(p,p',-q)\,,\\
{I^{(1,2)}_\pm(p,p',q)}^*&=&I^{(2,1)}_\mp(p,p',-q)\,,
\end{eqnarray}
indicate that only two types of integrals are independent. Therefore
we have nothing to lose if we restrict ourselves to study only  the
integrals $I^{(2,1)}_\pm$ and $I^{(2,2)}_\pm$ which are involved in
the structure of our amplitudes.

The next step is to evaluate these integrals by expanding them in sums of
integrals of $J$-functions as it results from Eqs. (\ref{h1}) and (\ref{h2}).
Thus we obtain
\begin{eqnarray}
I^{(2,1)}_\pm(p,p',q)&=&\frac{1}{\cosh^2
\pi\mu}\left\{A_\pm(p,p',q)+C_\pm(p,p',q)\right.\nonumber\\
&&~~~~~~~~~~~\left.-ie^{\mp\pi\mu}B_\pm(p,p',q)+ie^{\pm\pi\mu}B_\pm(p',p,q)\right\}\,,\\
I^{(2,2)}_\pm(p,p',q)&=&\frac{1}{\cosh^2
\pi\mu}\left\{e^{\mp 2\pi\mu}A_\pm(p,p',q)-C_\pm(p,p',q)\right.\nonumber\\
&&~~~~~~~~~~~\left.+ie^{\mp\pi\mu}[B_\pm(p,p',q)+B_\pm(p',p,q)]\right\}\,,
\end{eqnarray}
where the new integrals
\begin{eqnarray}
A_\pm(p,p',q)&=&\int_{0}^{\infty}ds\, s\, J_{\nu_\pm}(sp)J_{\nu_\pm}(sp')\,
e^{i q s}\nonumber\\
&=&\frac{i}{\pi}\frac{q}{(pp')^{\frac{3}{2}}}\frac{d}{dz}\,Q_{\pm
i\mu}[z-{\rm sign}(q)i0]\,,\label{Apm}\\
C_\pm(p,p',q)&=&\int_{0}^{\infty}ds\, s\, J_{-\nu_\pm}(sp)J_{-\nu_\pm}(sp')\,
e^{i q s}\nonumber\\
&=&\frac{i}{\pi}\frac{q}{(pp')^{\frac{3}{2}}}\frac{d}{dz}\,Q_{\mp
i\mu-1}[z-{\rm sign}(q)i0]\,,\label{Cpm}
\end{eqnarray}
can be calculated straightforwardly using Eq.(\ref{int1}) while the integrals
with indices of opposite signs,
\begin{eqnarray}
B_\pm(p,p',q)&=&\int_{0}^{\infty}ds\, s\, J_{\nu_\pm}(sp)J_{-\nu_\pm}(sp')\,
e^{i q s}=-\frac{(1\mp 2 i\mu)\cosh \pi\mu}{\pi
k^2(1+4\mu^2)}\left(\frac{p}{p'}\right)^{\frac{1}{2}\pm
i\mu}\nonumber\\
&\times&F_4\left(\frac{3}{2},1,\frac{3}{2}\pm i\mu,\frac{1}{2}\mp
i\mu;\frac{p^2}{k^2+{\rm sign}(q)i0},\frac{{p'}^2}{k^2+{\rm
sign}(q)i0}\right),\label{Bpm}
\end{eqnarray}
result from Eq. (\ref{int2}).

The functions $A_\pm$ and $C_\pm$ are expressed in terms of
Legendre functions of  second kind, $Q_\nu(z\pm i0)$, depending
on the variable
\begin{equation}\label{z}
z=\frac{p^2+{p'}^2-k^2}{2pp'}\,,
\end{equation}
which takes values in the domain $(-1, 1)$ because of the momentum
conservation in Eqs. (\ref{ampl1}) and (\ref{ampl2}).
Bearing in mind that the Legendre functions $Q_\nu$ have a branch
cut in this domain, we see that the small $\epsilon$ which assures
the convergence of these integrals determines the analytic form the
Legendre functions given in Appendix C. The functions $B_\pm$ have a
more complicated structure depending on the Appell hypergeometric
functions of double arguments $F_4$ \cite{GR}. Some technical
difficulties could arise here because of these functions, which are
less studied so far. Nevertheless, we have all the ingredients we
need for calculating the analytical expressions of these amplitudes.

The final form of the transition amplitude for the process $e^-\to e^- + \gamma$ will be:
\begin{eqnarray}\label{a1}
&&\langle (\vec{p}\,'\sigma')_-,
(\vec{k}\,\lambda)|{\bf S}_1|(\vec{p}\,\sigma)_-\rangle
=\delta^3(\vec{p}-{\vec{p}\,}'-\vec{k})\,\frac{ i e_0}{16\sqrt{\pi}}\left(\frac{p }{p'}\right)^{i\mu}\,\xi_{\sigma'}^+({\vec{p}\,}\,')
\,\vec{\sigma}\cdot
{\vec{\varepsilon}_{\lambda}(\vec{k})}^*\xi_{\sigma}({\vec{p}}\,)\nonumber\\
&&\times\{{\rm sign}(\sigma)\left[h_{\mu}(p,p',k)+l_{\mu}(p,p',k)\right]+
{\rm sign}(\sigma')\left[h_{-\mu}(p,p',k)+l_{-\mu}(p,p',k)\right]\}.
\end{eqnarray}
In the case $\gamma \to e^- + e^+$, the final expression of the transition amplitude is:
\begin{eqnarray}\label{a2}
&&\langle(\vec{p}\,\sigma)_-,(\vec{p}\,'\sigma')_+ |{\bf S}_1|(\vec{k}\,\lambda)
\rangle =-\delta^3(\vec{p}+\vec{p}\,'-\vec{k})\,
\frac{i e_0}{16\sqrt{\pi}}\left(\frac{p p'}{\omega^2}\right)^{i\mu}\,\xi_{\sigma}^+(\vec{p}\,)
\,\vec{\sigma}\cdot
\vec{\varepsilon}_{\lambda}(\vec{k})\,\eta_{\sigma'}({\vec{p}\,}')\nonumber\\
&&\times\{f_{\mu}(p,p',k)+g_{\mu}(p,p',k)-
{\rm sign}(\sigma \sigma')\left[f_{-\mu}(p,p',k)+g_{-\mu}(p,p',k)\right]\}.\nonumber\\
\end{eqnarray}
The newly introduced functions in (\ref{a1}),(\ref{a2}) $f_{\mu},g_{\mu},h_{\mu},l_{\mu}$ are:
\begin{eqnarray}\label{fu}
&&h_{\pm\mu}(p,p',k)=\mp\,\frac{\sqrt{k}}{2pp\,'}\frac{(-\mu^{2}\pm i\mu)}{\sinh(\pi\mu)\cosh^{2}(\pi\mu)}\left[\,_{2}F_{1}\left(1\mp i\mu,2\pm i\mu;2;\frac{1+z}{2}\right)\right.\nonumber\\
&&\left.\pm \sinh(\pi\mu)\,_{2}F_{1}\left(1\mp i\mu,2\pm i\mu;2;\frac{1-z}{2}\right)\right];\nonumber\\
&&l_{\pm\mu}(p,p',k)=i\,\sqrt{\frac{pp\,'}{k^5}}\frac{(1\mp 2i\mu)}{\pi(1+4\mu^2)\cosh(\pi\mu)}\nonumber\\
&&\times\left[e^{-\pi\mu}\left(\frac{p}{p\,'}\right)^{\frac{1}{2}\pm i\mu}\,F_{4}\left(\frac{3}{2},1,\frac{3}{2}\pm i\mu,\frac{1}{2}\mp i\mu;\frac{p^2}{k^2}-i0;\frac{p\,'^{2}}{k^2}-i0\right)\right.\nonumber\\
&&\left.-e^{\pi\mu}\left(\frac{p\,'}{p}\right)^{\frac{1}{2}\pm i\mu}\,F_{4}\left(\frac{3}{2},1,\frac{3}{2}\pm i\mu,\frac{1}{2}\mp i\mu;\frac{p\,'^{2}}{k^2}-i0;\frac{p^{2}}{k^2}-i0\right)\right];
\end{eqnarray}
\begin{eqnarray}\label{fu1}
&&f_{\pm\mu}(p,p',k)=\pm\,\frac{\sqrt{k}}{2pp\,'}\frac{(-\mu^{2}\pm i\mu)}{\sinh(\pi\mu)\cosh^{2}(\pi\mu)}\left[\cosh(2\pi\mu)\right.\nonumber\\
&&\left.\times_{2}F_{1}\left(1\mp i\mu,2\pm i\mu;2;\frac{1-z}{2}\right)\mp \sinh(\pi\mu)\,_{2}F_{1}\left(1\mp i\mu,2\pm i\mu;2;\frac{1+z}{2}\right)\right];\nonumber\\
&&g_{\pm\mu}(p,p',k)=-i\,\sqrt{\frac{pp\,'}{k^5}}\frac{(1\mp 2i\mu)}{\pi(1+4\mu^2)\cosh(\pi\mu)}\nonumber\\
&&\times\left[\left(\frac{p}{p\,'}\right)^{\frac{1}{2}\pm i\mu}F_{4}\left(\frac{3}{2},1,\frac{3}{2}\pm i\mu,\frac{1}{2}\mp i\mu;\frac{p^2}{k^2}+i0;\frac{p\,'^{2}}{k^2}+i0\right)\right.\nonumber\\
&&\left.+\left(\frac{p\,'}{p}\right)^{\frac{1}{2}\pm i\mu}\,F_{4}\left(\frac{3}{2},1,\frac{3}{2}\pm i\mu,\frac{1}{2}\mp i\mu;\frac{p\,'^{2}}{k^2}+i0;\frac{p^{2}}{k^2}+i0\right)\right].
\end{eqnarray}

We observe that our amplitudes depends on the parameter $\mu=m/\omega$. This dependence encode the influence of the space expansion on the particle production process.  Further we plot the real and imaginary parts of the functions defined in (\ref{fu}), (\ref{fu1}) as function of parameter $\mu$ for different values of the momenta $p,\,p\,' ,\,k$. The Appell functions $F_{4}$ are less studied and for that reason we will approximate our functions $l_{\mu},g_{\mu}$, observing that both are proportional with a factor $[(1+4\mu^2)\cosh(\pi\mu)]^{-1}$, which makes these functions convergent for large $\mu$. So the relevant contributions for the graphs of the functions $l_{\mu},\,g_{\mu}$ will come from this factor. This can be seen by approximating the Appell functions with hypergeometric Gauss functions and then plotting the result. In our further considerations, the Appell functions $F_{4}$ from $l_{\mu},\,g_{\mu}$, will be taken to be equal to unity . In our graphs we introduce the notation $\mu=u=\frac{m}{\omega}$ .

\begin{figure}[H] \centering
\includegraphics[scale=0.5]{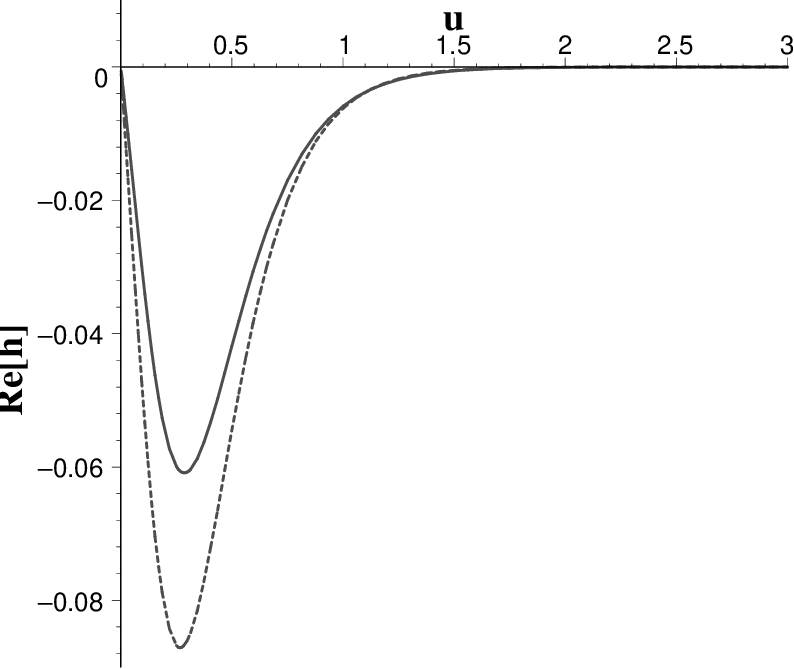}
\caption{The real part of $h_{\mu}$ as a function of $u$. The
solid line is for $z=0.5$ and the dashed line for $z=0.9$.}
\label{f1}
\end{figure}

\begin{figure}[H]\centering
\includegraphics[scale=0.5]{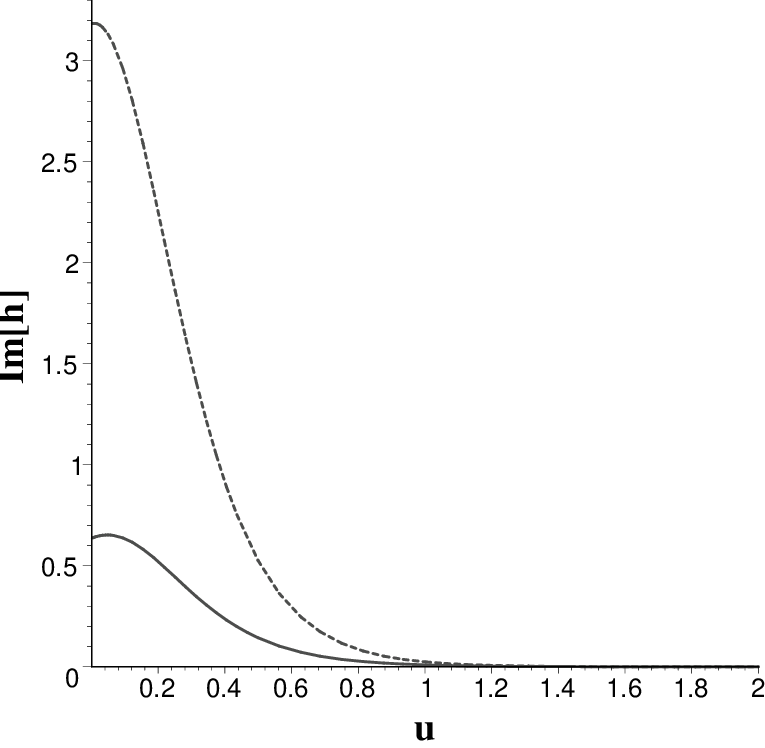}
\caption{The imaginary part of $h_{\mu}$ as a function of $u$. The
solid line is for $z=0.5$ and the dashed line for $z=0.9$.}
\label{f2}
\end{figure}

\begin{figure}[H]\centering
\includegraphics[scale=0.5]{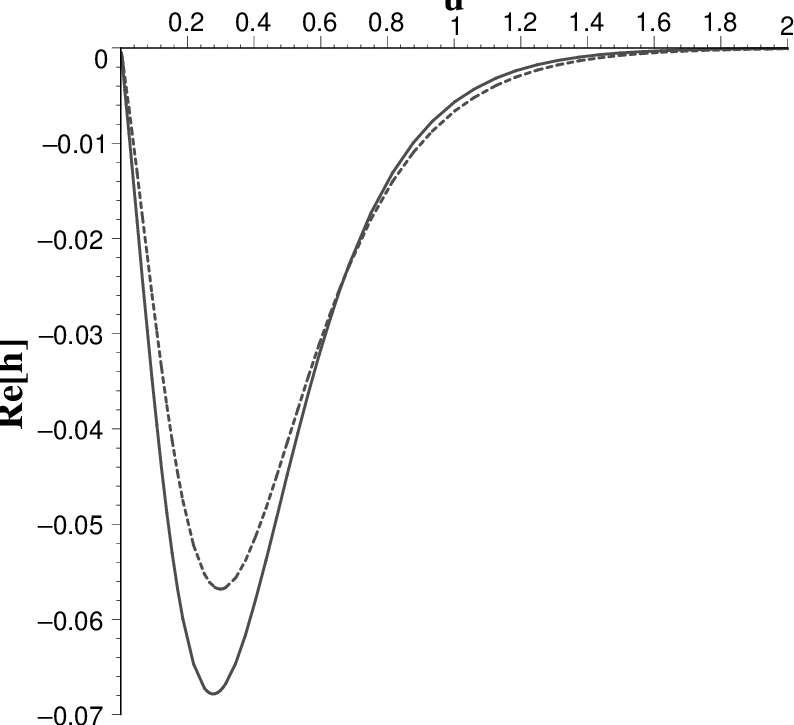}
\caption{The real part of $h_{\mu}$ as a function of $u$. The
solid line is for $z=0.7$ and the dashed line for $z=0.2$.}
\label{f3}
\end{figure}

\begin{figure}[H]\centering
\includegraphics[scale=0.5]{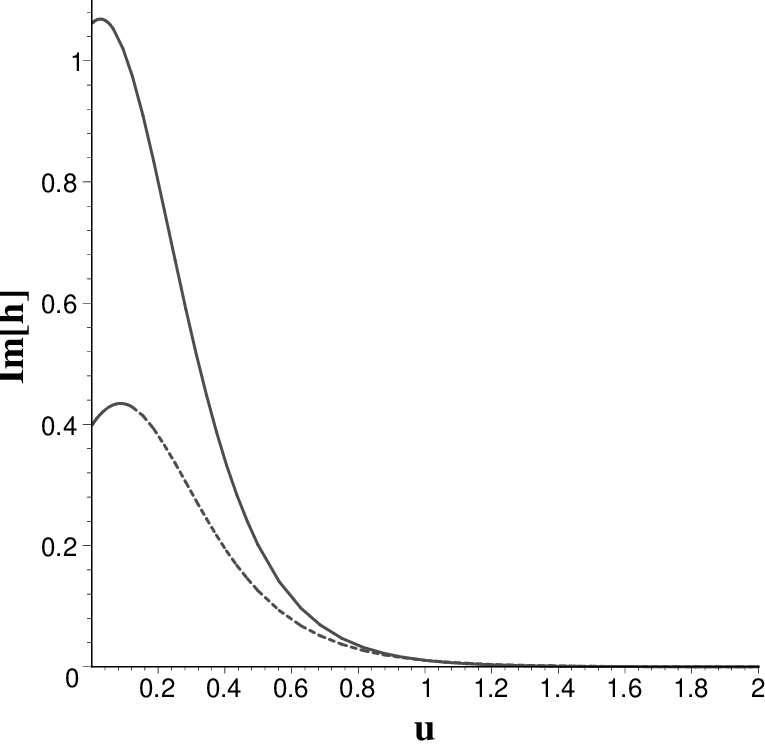}
\caption{The imaginary part of $h_{\mu}$ as a function of $u$. The
solid line is for $z=0.7$ and the dashed line for $z=0.2$.}
\label{f4}
\end{figure}
As we can observe from Figs.(\ref{f1})-(\ref{f4}), the real and imaginary part of the function $h_{\mu}$ are finite in origin and converge rapidly to zero for large values of the parameter $\mu$.

\begin{figure}[H]\centering
\includegraphics[scale=0.5]{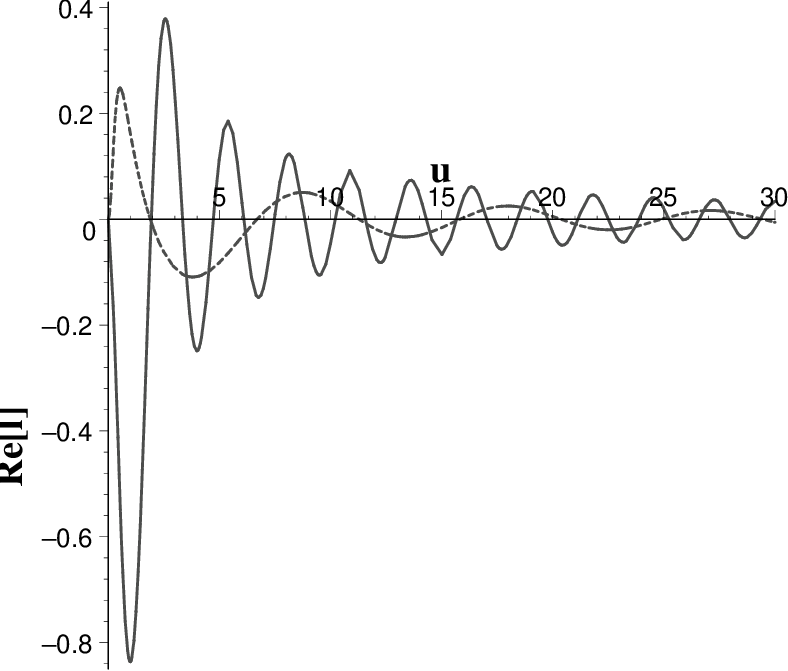}
\caption{The real part of $l_{\mu}$ as a function of $u$. The
solid line is for $\frac{p}{p\,'}=\frac{1}{10}$ and the dashed line for $\frac{p}{p\,'}=\frac{1}{2}$.}
\label{f5}
\end{figure}

\begin{figure}[H]\centering
\includegraphics[scale=0.5]{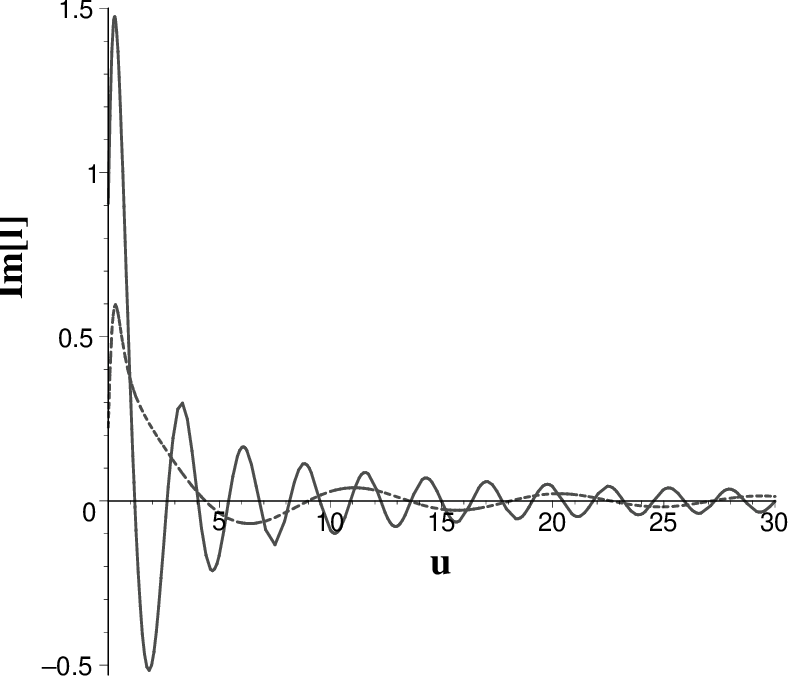}
\caption{The imaginary part of $l_{\mu}$ as a function of $u$. The
solid line is for $\frac{p}{p\,'}=\frac{1}{10}$ and the dashed line for $z=\frac{p}{p\,'}=\frac{1}{2}$.}
\label{f6}
\end{figure}
In the case of the function $l_{\mu}$, both the real and imaginary parts are finite in origin and converge, our graphical analysis showing that the oscillatory behaviour of these functions approaches zero for $\mu=200$. This oscillatory behaviour is given by the exponential factors $e^{\pm \pi \mu}$ combined with the oscillatory factors $\left(\frac{p}{p\,'}\right)^{\frac{1}{2}\pm i\mu}$. From  Figs.(\ref{f1})-(\ref{f6})) we can conclude that our functions that define the amplitude of the process in which the electron emits one photon are convergent.

Let us see what happens in the case of pair production by a single photon.
\begin{figure}[H]\centering
\includegraphics[scale=0.5]{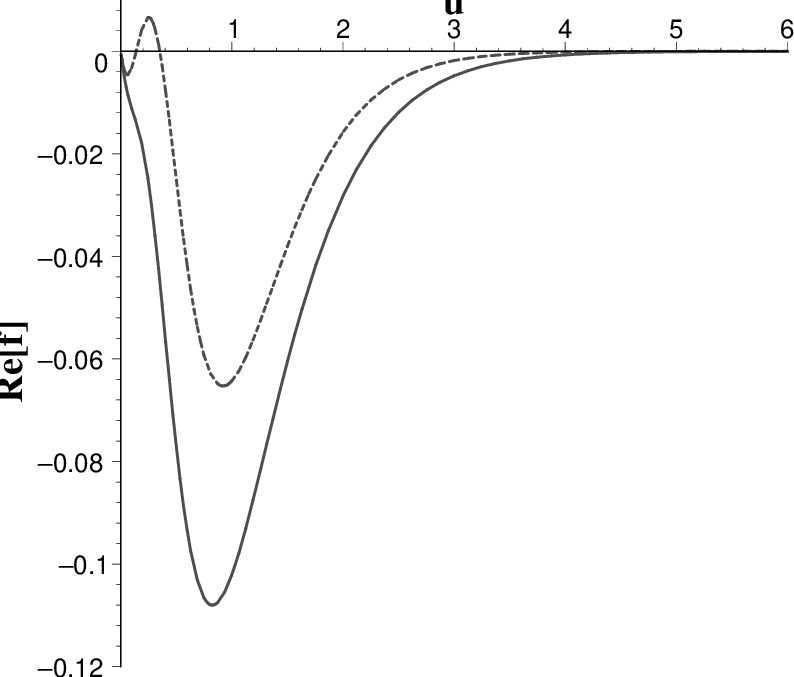}
\caption{The real part of $f_{\mu}$ as a function of $u$. The
solid line is for $z=0.5$ and the dashed line for $z=0.9$.}
\label{f7}
\end{figure}

\begin{figure}[H]\centering
\includegraphics[scale=0.5]{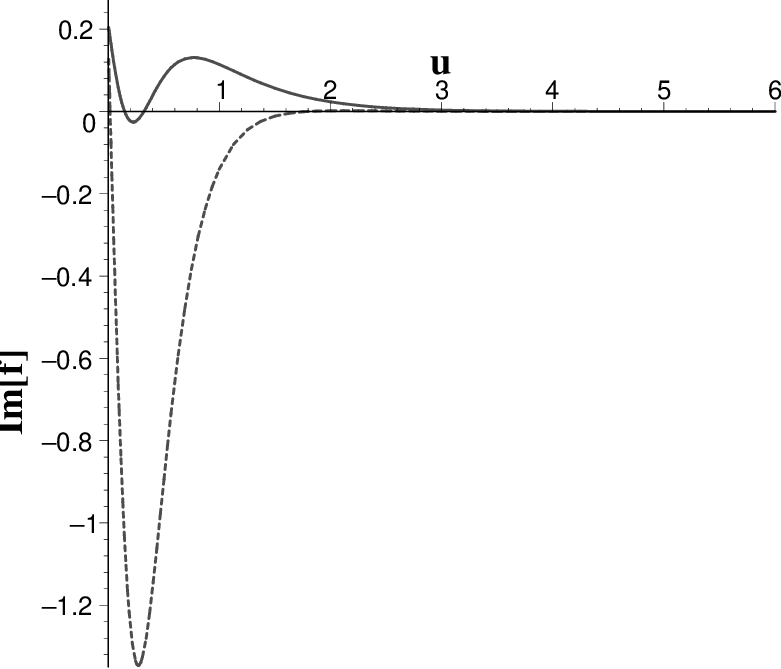}
\caption{The imaginary part of $f_{\mu}$ as a function of $u$. The
solid line is for $z=0.5$ and the dashed line for $z=0.9$.}
\label{f8}
\end{figure}

\begin{figure}[H]\centering
\includegraphics[scale=0.5]{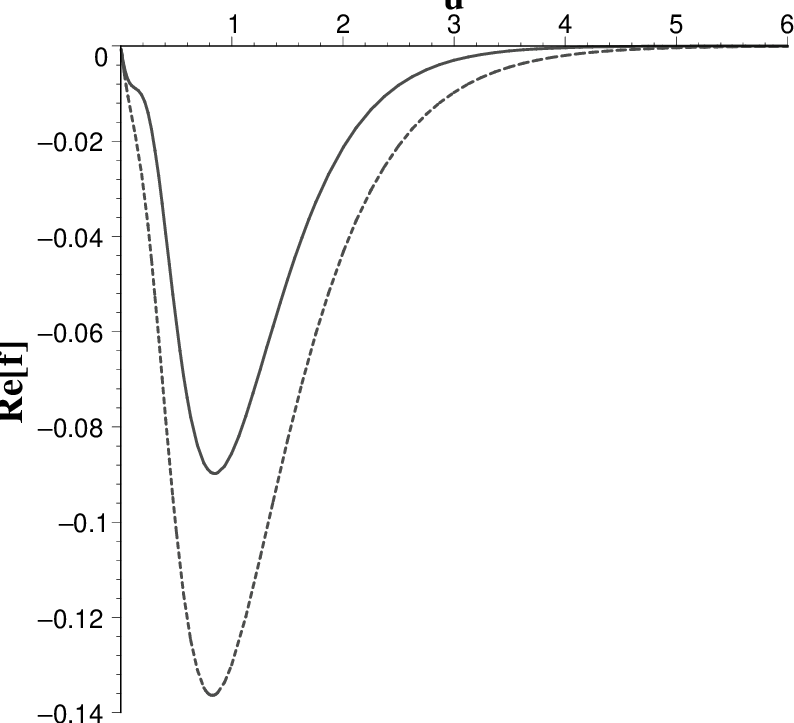}
\caption{The real part of $f_{\mu}$ as a function of $u$. The
solid line is for $z=0.7$ and the dashed line for $z=0.2$.}
\label{f9}
\end{figure}

\begin{figure}[H]\centering
\includegraphics[scale=0.5]{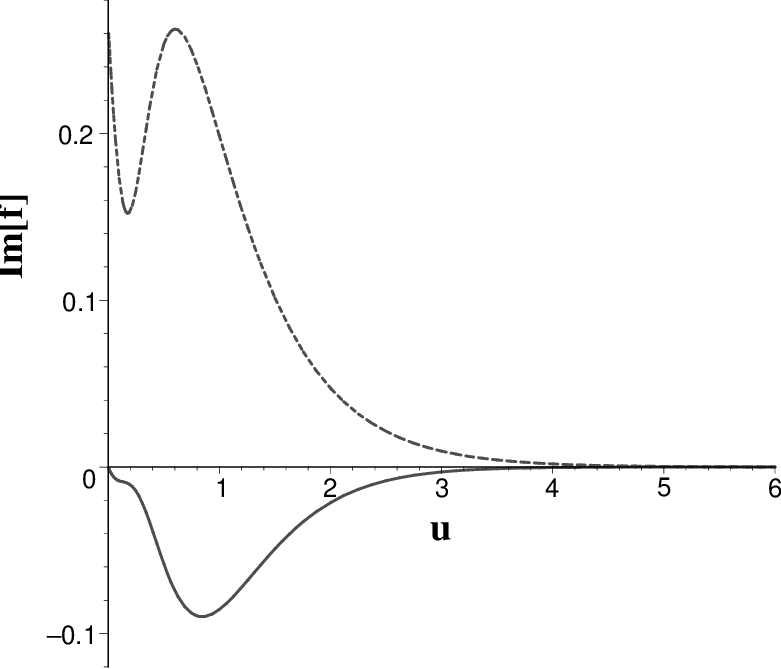}
\caption{The imaginary part of $f_{\mu}$ as a function of $u$. The
solid line is for $z=0.7$ and the dashed line for $z=0.2$.}
\label{f10}
\end{figure}

\begin{figure}[H]\centering
\includegraphics[scale=0.5]{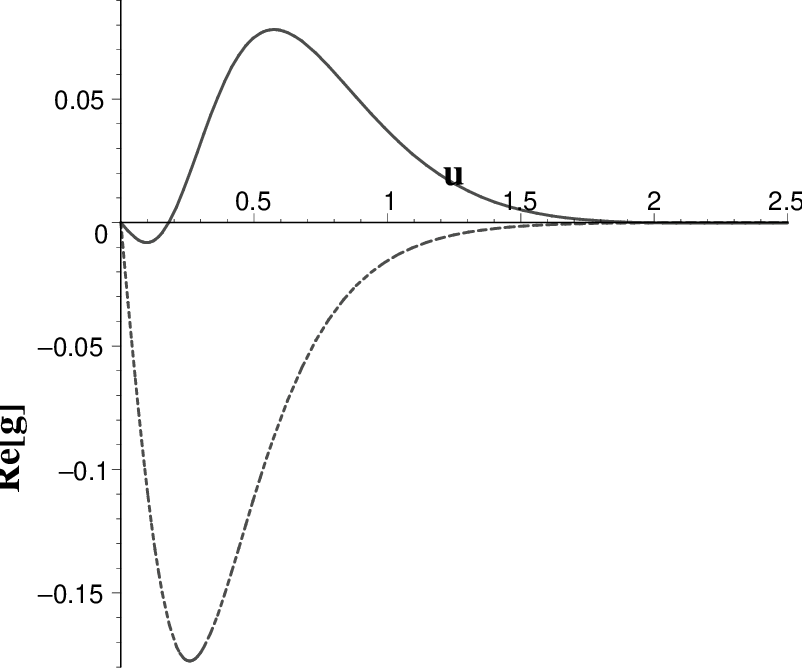}
\caption{The real part of $g_{\mu}$ as a function of $u$. The
solid line is for $\frac{p}{p\,'}=\frac{1}{10}$ and the dashed line for $\frac{p}{p\,'}=\frac{1}{2}$.}
\label{f11}
\end{figure}

\begin{figure}[H]\centering
\includegraphics[scale=0.5]{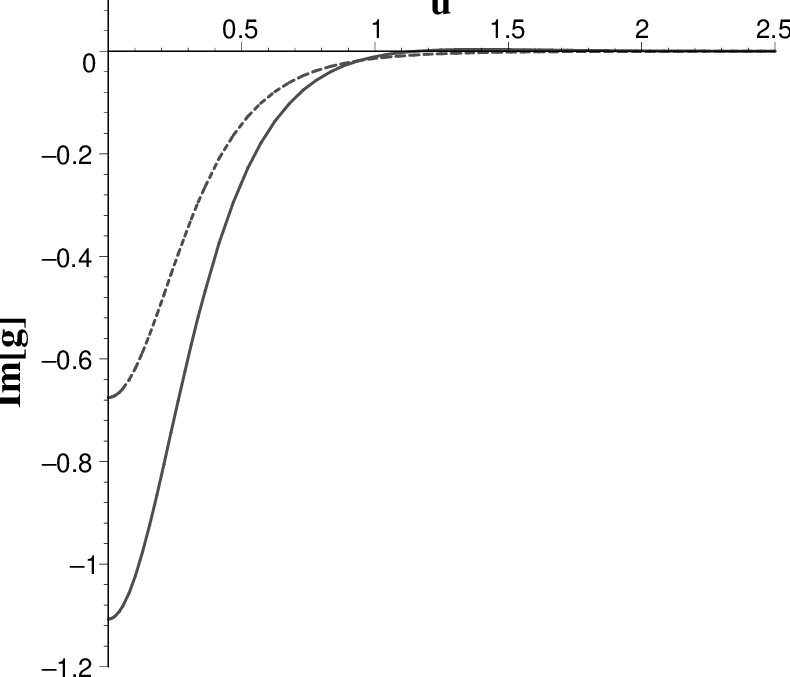}
\caption{The imaginary part of $g_{\mu}$ as a function of $u$. The
solid line is for $\frac{p}{p\,'}=\frac{1}{10}$ and the dashed line for $\frac{p}{p\,'}=\frac{1}{2}$.}
\label{f12}
\end{figure}
As in the case presented above, our functions $f_{\mu},g_{\mu}$ are very convergent and finite in origin. From all the above plots it is obviously that our result for the two amplitudes presented in this section is valid in the sense that our functions are very convergent. In the next sections we will make a more detailed analysis of their square modulus for defining the probability of the pair production in the presence of interactions.

Also we must specify that the phase factors $\left(\frac{p p'}{\omega^2}\right)^{i\mu}\,\,,\left(\frac{p }{p'}\right)^{i\mu}$ will give no contributions to our probabilities and for that reason they are not included in our functions defined in (\ref{fu}) and (\ref{fu1}).

\subsection{Electromagnetic particle creation}

The examples we analyze now in more detail are the related amplitudes of  the transitions $vac \to e^- + e^+ +\gamma$ and $e^- + e^+ +\gamma \to
vac$ that read
\begin{eqnarray}
&&\langle (\vec{p}\,\sigma)_-, (\vec{p}\,'\sigma')_+,(\vec{k}\,\lambda) |{\bf S}_1|0\rangle=-\langle 0|{\bf S}_1| (\vec{p}\,\sigma)_-, (\vec{p}\,'\sigma')_+,(\vec{k}\,\lambda) \rangle ^*\nonumber\\
&&\hspace*{18mm}=-\delta^3(\vec{p}+\vec{p}\,'+\vec{k})\,
\frac{i e_0}{16\sqrt{\pi}}\sqrt{\frac{p p'}{k}}\,
\left(\frac{p p'}{\omega^2}\right)^{i\mu}\,\xi_{\sigma}^+(\vec{p}\,)
\,\vec{\sigma}\cdot
\vec{\varepsilon}_{\lambda}(\vec{k})^*\,\eta_{\sigma'}({\vec{p}\,}')\nonumber\\
&&\hspace*{22mm}\times\left[e^{\pi\mu} I^{(2,2)}_+(p,p',-k)-{\rm
sign}(\sigma\sigma')\,e^{-\pi\mu}I^{(2,2)}_-(p,p',-k)\right]\label{ampl2}\,,\end{eqnarray}
The final form of this amplitude reads:
\begin{eqnarray}\label{as}
&&\langle (\vec{p}\,\sigma)_-, (\vec{p}\,'\sigma')_+,(\vec{k}\,\lambda) |{\bf S}_1|0\rangle=
-\delta^3(\vec{p}+\vec{p}\,'+\vec{k})\,
\frac{i e_0}{16\sqrt{\pi}}
\left(\frac{p p'}{\omega^2}\right)^{i\mu}\,\xi_{\sigma}^+(\vec{p}\,)\,\vec{\sigma}\cdot
\vec{\varepsilon}_{\lambda}(\vec{k})^*\,\eta_{\sigma'}({\vec{p}\,}')\nonumber\\
&&\times\{n_{\mu}(p,p',k)+g_{\mu}(p,p',k)-
{\rm sign}(\sigma\sigma')\left[n_{-\mu}(p,p',k)+g_{-\mu}(p,p',k)\right]\},
\end{eqnarray}
where the functions $g_{\mu}$ are defined in (\ref{fu}) and the newly introduced functions $n_{\mu}$ are defined as follows:
\begin{eqnarray}
n_{\pm\mu}(p,p',k)=\mp\,\frac{\sqrt{k}}{2pp\,'}\frac{(-\mu^{2}\pm i\mu)}{\sinh(\pi\mu)\cosh^{2}(\pi\mu)}\left[\,_{2}F_{1}\left(1\mp i\mu,2\pm i\mu;2;\frac{1-z}{2}\right)\right.\nonumber\\
\left.\mp \sinh(\pi\mu)\,_{2}F_{1}\left(1\mp i\mu,2\pm i\mu;2;\frac{1+z}{2}\right)\right].
\end{eqnarray}
Plotting the new function $n_{\mu}$ that enters in our amplitude, we obtain:

\begin{figure}[H]\centering
\includegraphics[scale=0.5]{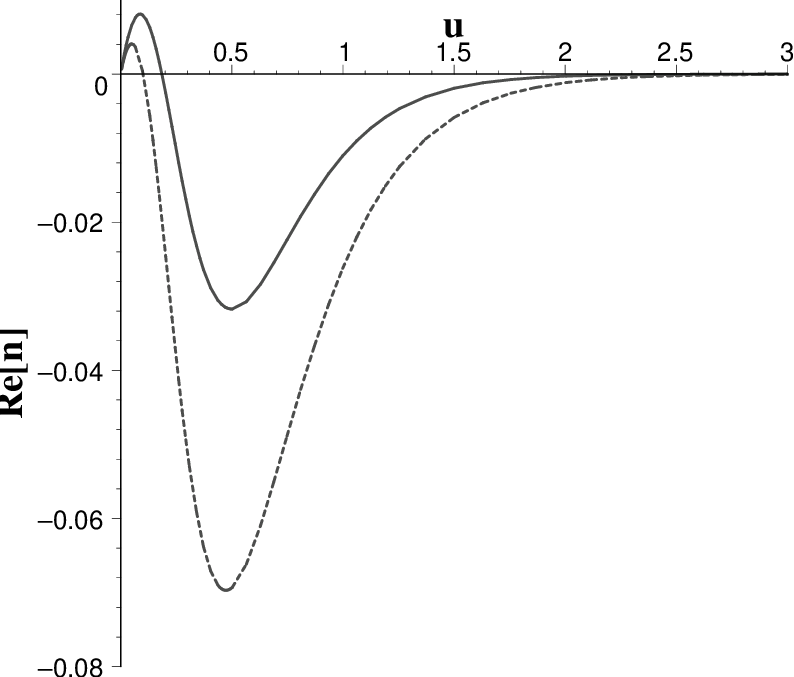}
\caption{The real part of $n_{\mu}$ as a function of $u$. The
solid line is for $z=0.5$ and the dashed line for $z=0.9$.}
\label{f13}
\end{figure}

\begin{figure}[H]\centering
\includegraphics[scale=0.5]{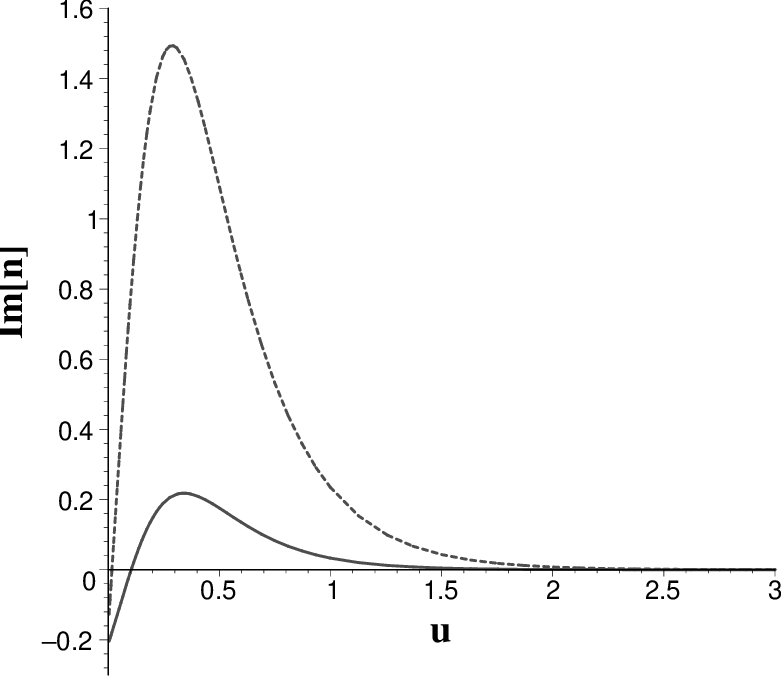}
\caption{The imaginary part of $n_{\mu}$ as a function of $u$. The
solid line is for $z=0.5$ and the dashed line for $z=0.9$.}
\label{f14}
\end{figure}

\begin{figure}[H]\centering
\includegraphics[scale=0.5]{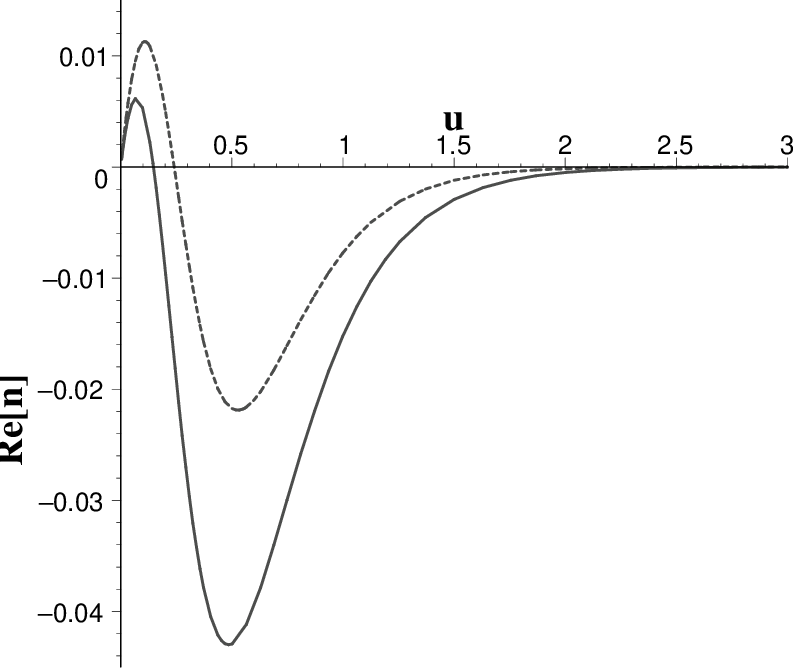}
\caption{The real part of $n_{\mu}$ as a function of $u$. The
solid line is for $z=0.7$ and the dashed line for $z=0.2$.}
\label{f15}
\end{figure}

\begin{figure}[H]\centering
\includegraphics[scale=0.5]{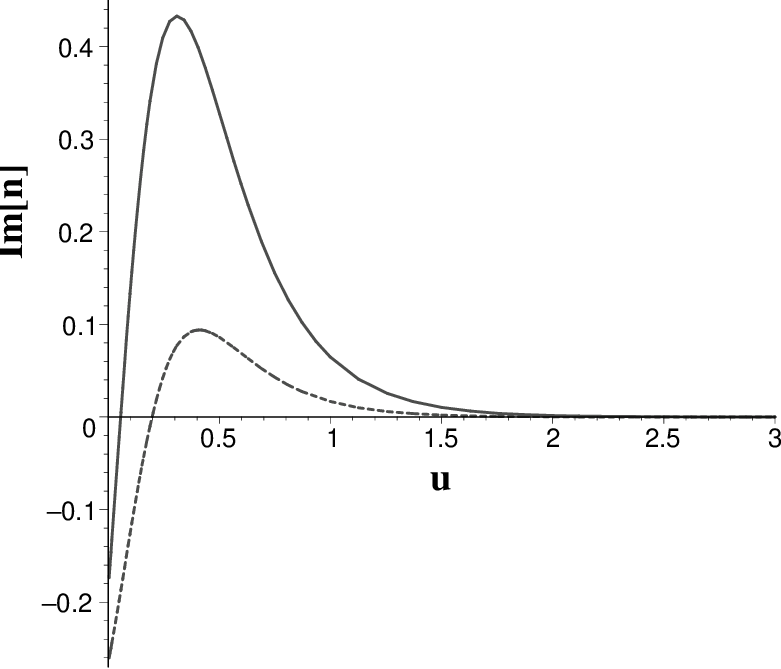}
\caption{The imaginary part of $n_{\mu}$ as a function of $u$. The
solid line is for $z=0.7$ and the dashed line for $z=0.2$.}
\label{f16}
\end{figure}
For function $g_{\mu}$ the graphs are presented in the previous section. The real and imaginary parts of the function $n_{\mu}$ are also convergent and finite in origin.

Simple kinetic parameters can be introduced in the orthogonal local
frame $\{\vec{e}_i\}$ where $\vec{k}=k\vec{e}_3$. In this frame we
take the electron and positron momenta in the plane $(1,3)$ denoting
their spherical coordinates as $\vec{p}=(p,\alpha,0)$ and
${\vec{p}\,}'=(p',\beta,\pi)$ where $\alpha,\, \beta\in(0,\pi)$.
Then the momentum conservation gives the equations
$k=p\cos\alpha+p'\cos\beta$ and $p\sin\alpha=p'\sin\beta$ from which
we deduce
\begin{equation}\label{ppalphabeta}
\frac{p}{k}=\frac{\sin\beta}{\sin(\alpha+\beta)}\,,\quad
\frac{p'}{k}=\frac{\sin\alpha}{\sin(\alpha+\beta)}\,.
\end{equation}
Moreover, from Eq. (\ref{z}) we obtain $z=-\cos(\alpha+\beta)$ since
the angle between $\vec{p}$ and ${\vec{p}\,}'$ is just
$\alpha+\beta$.

Using now  Eqs. (\ref{Apm}), (\ref{Cpm}), (\ref{Bpm}) and (\ref{ppalphabeta}) and taking $q=-k$ we obtain the final expression
\begin{eqnarray}\label{AAA}
&&\langle (\vec{p}\,\sigma)_-, (\vec{p}\,'\sigma')_+,(\vec{k}\,\lambda) |{\bf S}_1|0\rangle
=\\
&&-\delta^3(\vec{p}+{\vec{p}\,}'+\vec{k})\,\frac{i e_0}{16\sqrt{\pi}}\,\frac{1}{k^{\frac{3}{2}}}\,\left(\frac{k}{\omega}\right)^{2i\mu}\left[\frac{\sin\alpha\sin\beta}{\sin^2(\alpha+\beta)}\right]^{i\mu}
M_{\sigma,\sigma'}(\lambda)\nonumber\\
&&\times \left\{{\cal F}_\mu(\alpha,\beta)+{\cal G}_\mu(\alpha,\beta)-{\rm
sign}(\sigma\sigma')\left[{\cal F}_{-\mu}(\alpha,\beta)+{\cal
G}_{-\mu}(\alpha,\beta)\right]\right\}\,,
\end{eqnarray}
where the functions
\begin{eqnarray}
{\cal F}_\mu(\alpha,\beta)
&=&\frac{\mu(\mu-i)}{2\cosh^2\pi\mu\,\sinh\pi\mu}\,\frac{\sin^2(\alpha+\beta)}
{\sin\alpha\sin\beta}
\left[F\left(1-i\mu,2+i\mu;2;\cos^2\frac{\alpha+\beta}{2}\right)\right.\nonumber\\
&&~~~~~\left. -\sinh\pi\mu\,
F\left(1-i\mu,2+i\mu;2;\sin^2\frac{\alpha+\beta}{2}\right)\right]\\
{\cal
G}_\mu(\alpha,\beta)&=&\,\frac{2\mu-i}{\pi(1+4\mu^2)}\,\frac{1}{\cosh\pi\mu}\,
\frac{\sqrt{\sin\alpha\sin\beta}}{\sin(\alpha+\beta)}
\left[\left(\frac{\sin\alpha} {\sin\beta}\right)^{\frac{1}{2}+i\mu}\right.\nonumber\\
&&~~\left.\times F_4\left(\frac{3}{2},1,\frac{3}{2}+i\mu,\frac{1}{2}-i\mu;
\frac{\sin^2\alpha}{\sin^2(\alpha+\beta)}+i0,\frac{\sin^2\beta}{\sin^2(\alpha+\beta)}
+i0\right)\right.\nonumber\\
&&\hspace{62mm}\left.+\, (\alpha \longleftrightarrow \beta)\right]
\end{eqnarray}
depend only on the angles $\alpha$ and $\beta$ and the parameter $\mu$. The matrix elements $M_{\sigma,\sigma'}(\lambda)$ are given in Appendix A.

It is remarkable that the above amplitudes depend on the fermion mass and the external gravity only through the parameter $\mu=\frac{m}{\omega}$. This  becomes very small under inflation when the Hubble constant $\omega$ is extremely large. This situation is well approximated by the limit of the amplitude (\ref{AAA}) for $\mu\to 0$. Taking into account that in this limit (when $\nu_{\pm}\to \frac{1}{2}$) the Hankel functions are of the
form (\ref{Ap1}) we can evaluate the integral
\begin{equation}
\lim_{\mu\to 0} I^{(2,2)}_{\pm}(p,p',-k)=\frac{2
i}{\pi}\frac{1}{\sqrt{p p'}}\frac{1}{k+p+p'-i0}\,,
\end{equation}
which leads to the amplitudes
\begin{eqnarray}\label{AAAlim}
&&\lim_{\mu\to 0} \langle (\vec{p}\,\sigma)_-, (\vec{p}\,'\sigma')_+,(\vec{k}\,\lambda) |{\bf S}_1|0\rangle \nonumber\\
&&\hspace*{20mm}=\frac{e_0}{8(\pi
k)^{\frac{3}{2}}}\,\delta^3(\vec{p}+{\vec{p}\,}'+\vec{k})
\delta_{\sigma,-\sigma'} M_{\sigma,-\sigma}(\lambda)
\frac{\cos\frac{\alpha+\beta}{2}}{\cos\frac{\alpha}{2}\cos\frac{\beta}{2}}\,.
\end{eqnarray}

These amplitudes are non-vanishing only if $\sigma'=-\sigma$. Thus,
for $\lambda=1$ we find two non-vanishing amplitudes proportional to
\begin{equation}
\frac{e_0}{2(2 \pi k)^{\frac{3}{2}}}\cos\frac{\alpha+\beta}{2}\times
\left\{
\begin{array}{cll}
\tan\frac{\alpha}{2}&{\rm for}&\sigma=-\sigma'=-\frac{1}{2}\\
(-\tan\frac{\beta}{2})&{\rm for}&\sigma=-\sigma'=\frac{1}{2}
\end{array}\right.
\end{equation}
Similar results written for $\lambda=-1$ show that all these
amplitudes vanishes for $\alpha=\beta=0$ when $e^-$ and $e^+$ have
parallel momenta in the same direction. However, whether $e^-$ and
$e^+$ have parallel momenta, but in opposite directions, i. e.
$(\alpha=\pi,\,\beta=0)$ or $(\alpha=0,\,\beta=\pi)$, we can not use
the general formula (\ref{AAA}) being forced to reconsider the
momentum conservation.

Let us take, for example, the {\em out} state
with a photon having $\vec{k}=-k\,\vec{e}_3$ and $\lambda=1$, an
electron of parameters $\vec{p}=p\,\vec{e}_3$ and
$\sigma=\frac{1}{2}$ and a positron with
${\vec{p}\,}'=(k-p)\vec{e}_3$ provided $p>k$ and
$\sigma'=-\frac{1}{2}$. Then the resulting amplitude for $\mu\to 0$
reads
\begin{equation}
\lim_{\mu\to 0}  \langle (\vec{p}\,\textstyle{\frac{1}{2}})_-, (\vec{p}\,'\textstyle{-\frac{1}{2}})_+,(\vec{k}\, 1) |{\bf S}_1|0\rangle=\frac{\textstyle e_0}{\textstyle 2(2 \pi )^{\frac{3}{2}}}\frac{\textstyle 1}{\textstyle p\sqrt{k}}\, \delta^3(\vec{p}+{\vec{p}\,}'+\vec{k})\,.
\end{equation}
The conclusion is that under inflation the effect of electromagnetic particle
creation is favoured only when it produces pairs of fermions moving
in opposite directions.
This phenomenon  could
be one of the mechanisms of separating the matter and antimatter
between themselves.

\section{Transition probabilities}

In this section we explore the physical consequences of our calculations. Because we use here the methods based on perturbations the outcome of our calculations are the probabilities of transitions. We will study in detail the properties of our probabilities paying a special attention to the limit of large expansion factor comparatively with the particle mass. All the three processes analysed here have their amplitudes proportional with a $\delta^3(\vec{p}\,)$ function.By squaring the amplitudes we will obtain terms of the type $|\delta^3(\vec{p}\,)|^2=V\,\delta^3(\vec{p}\,)$, and we can define in this way the probability per unit of volume. In this section we will analyse only the probability transitions for $vac \to e^- + e^+ +\gamma$ and $\gamma \to e^- + e^+ $.

For the electron-positron pair production by a single photon the probability in volume unit is obtained by squaring the amplitude and summing after final helicities $\sigma,\,\sigma'$:
\begin{eqnarray}\label{pr1}
&&\mathcal{P}_{\gamma \rightarrow e^-e^+}=\frac{1}{2}\sum_{\sigma\sigma'}\frac{|\langle(\vec{p}\,\sigma)_-,(\vec{p}\,'\sigma')_+ |{\bf S}_1|(\vec{k}\,\lambda)
\rangle|^2}{V} =\frac{ e^{2}_{0}}{256\pi}\delta^3(\vec{p}+\vec{p}\,'-\vec{k})\,
|\xi_{\sigma}^+(\vec{p}\,)
\,\vec{\sigma}\cdot\vec{\varepsilon}_{\lambda}(\vec{k})\,\eta_{\sigma'}({\vec{p}\,}')|^2\nonumber\\
&&\times\left[|f_{\mu}(p,p',k)|^2+|g_{\mu}(p,p',k)|^2+|f_{-\mu}(p,p',k)|^2+|g_{-\mu}(p,p',k)|^2+f_{\mu}(p,p',k)g^*_{\mu}(p,p',k)\right.\nonumber\\
&&\left.+f^*_{\mu}(p,p',k)g_{\mu}(p,p',k)+f^*_{-\mu}(p,p',k)g_{-\mu}(p,p',k)+f_{-\mu}(p,p',k)g^*_{-\mu}(p,p',k)\right.\nonumber\\
&&\left.-{\rm sign}(\sigma\sigma')(f^*_{\mu}(p,p',k)f_{-\mu}(p,p',k)+f^*_{-\mu}(p,p',k)f_{\mu}(p,p',k)+g^*_{-\mu}(p,p',k)g_{\mu}(p,p',k))\right.\nonumber\\
&&\left.-{\rm sign}(\sigma\sigma')(g^*_{\mu}(p,p',k)g_{-\mu}(p,p',k)+f^*_{-\mu}(p,p',k)g_{\mu}(p,p',k)+f_{-\mu}(p,p',k)g^*_{\mu}(p,p',k))\right.\nonumber\\
&&\left.-{\rm sign}(\sigma\sigma')(g^*_{-\mu}(p,p',k)f_{\mu}(p,p',k)+f^*_{\mu}(p,p',k)g_{-\mu}(p,p',k))\right].
\end{eqnarray}
For obtaining the total probability in volume unit we must integrate (\ref{pr1}) after the final momenta $p,\,p\,'$, $\mathcal{P}^{tot}_{\gamma\rightarrow e^-e^+}=\int \mathcal{P}_{\gamma\rightarrow e^-e^+}\,\frac{d^3p}{(2\pi)^{3}}\frac{d^3p\,'}{(2\pi)^{3}}$. The integrals after the final momenta are very complicated and we restrict the analysis only to the probability given in (\ref{pr1}). As we can observe from (\ref{pr1}), there are nonvanishing probabilities for pair production in the both helicity conserving/nonconserving cases. Because our probability equation is complicated, only a graphical analysis will help us to understand better the physics beyond this formula. Plotting the probability (\ref{pr1}) as function of parameter $u=m/\omega$ for different values of the parameter $z$ we obtain:
\begin{figure}[H]\centering
\includegraphics[scale=0.4]{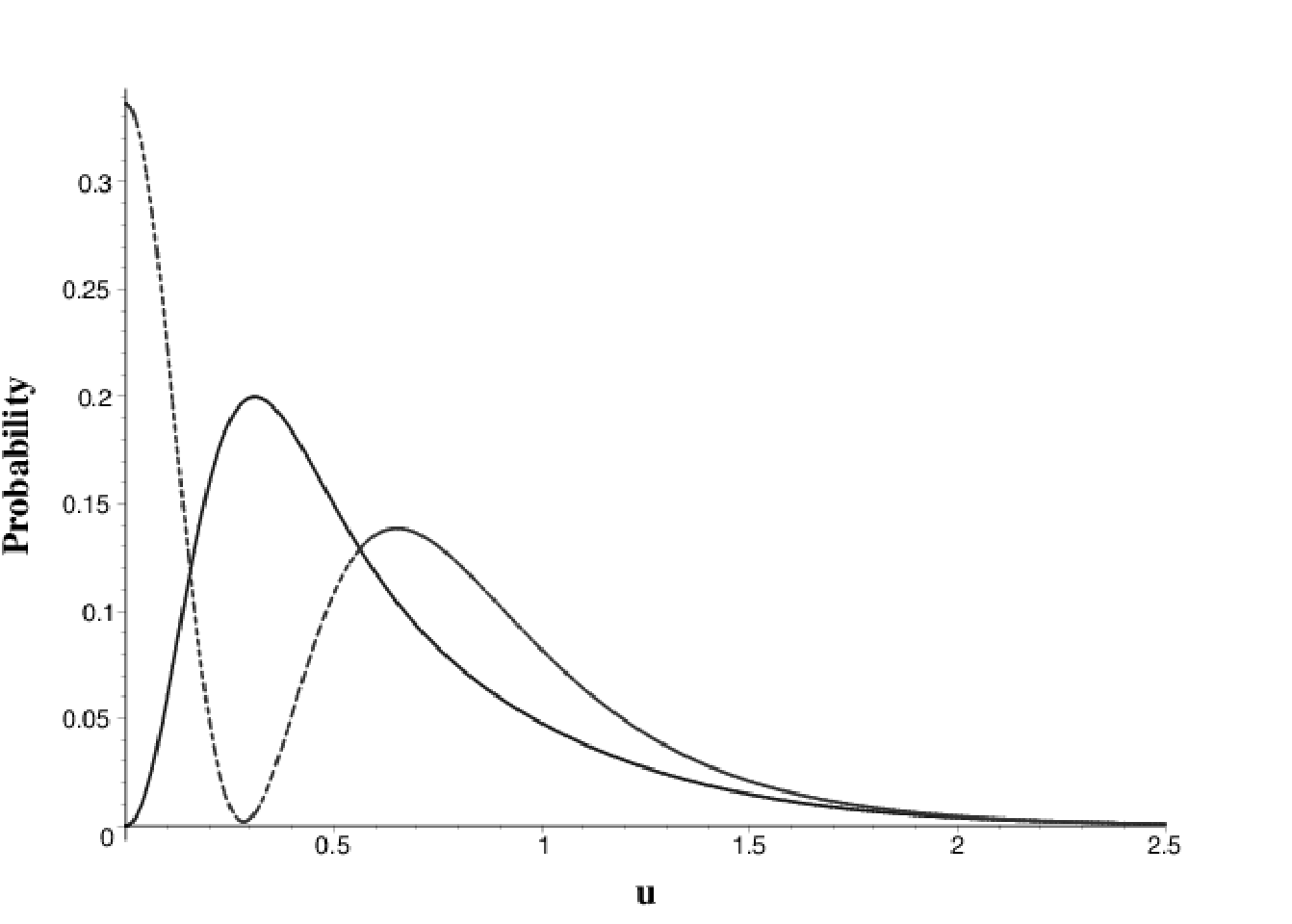}
\caption{$\mathcal{P}_{\gamma \rightarrow e^-e^+}$ as a function of $u$ for $z=0.2$ and $\frac{p}{p\,'}=\frac{1}{2}$. The dashed line represents the case of helicity conservation and the solid line the case when helicity is not conserved.}
\label{f17}
\end{figure}

\begin{figure}[H]\centering
\includegraphics[scale=0.4]{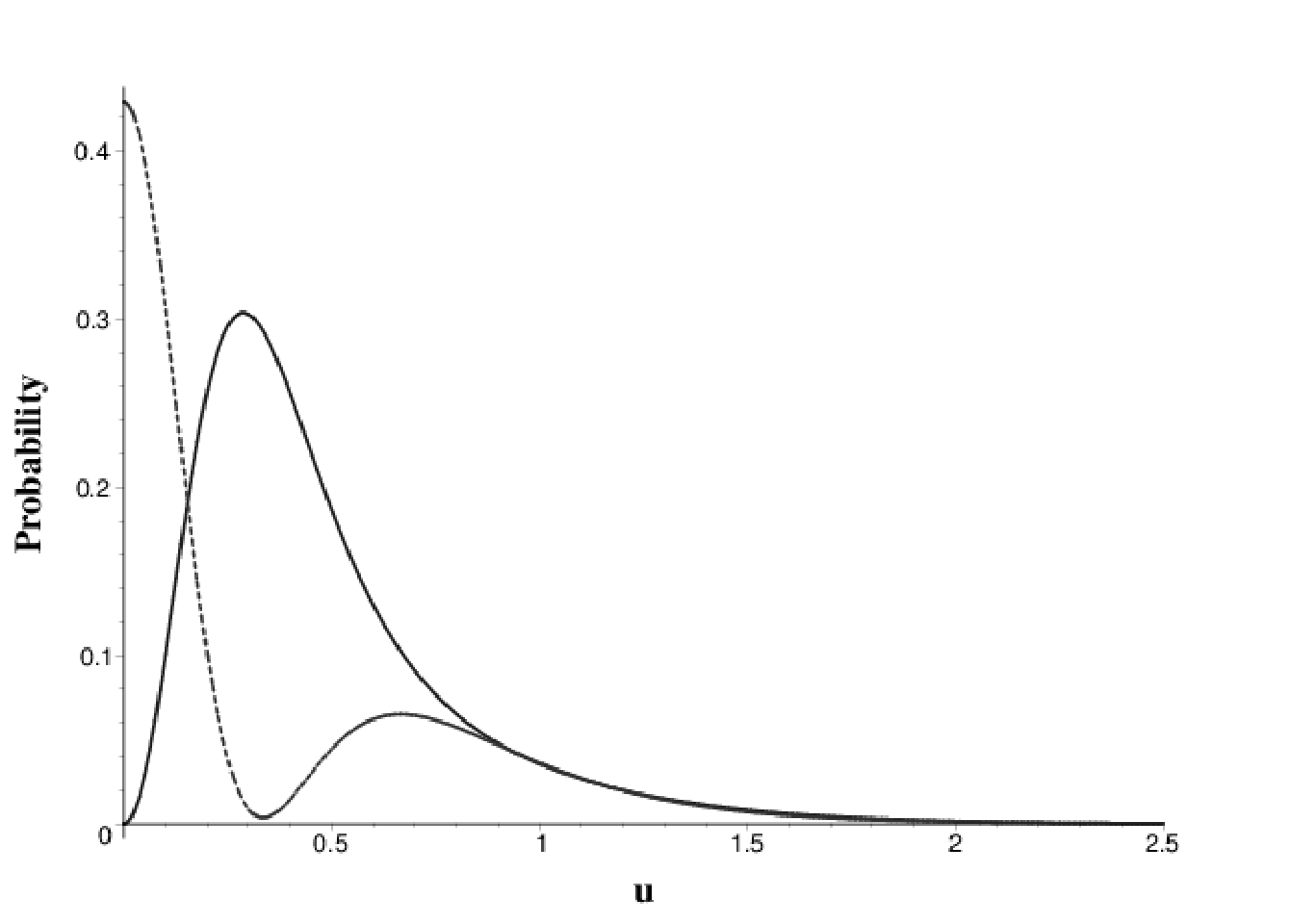}
\caption{$\mathcal{P}_{\gamma \rightarrow e^-e^+}$ as a function of $u$ for $z=0.5$ and $\frac{p}{p\,'}=\frac{1}{2}$. The dashed line represents the case of helicity conservation and the solid line the case when helicity is not conserved.}
\label{f18}
\end{figure}

\begin{figure}[H]\centering
\includegraphics[scale=0.4]{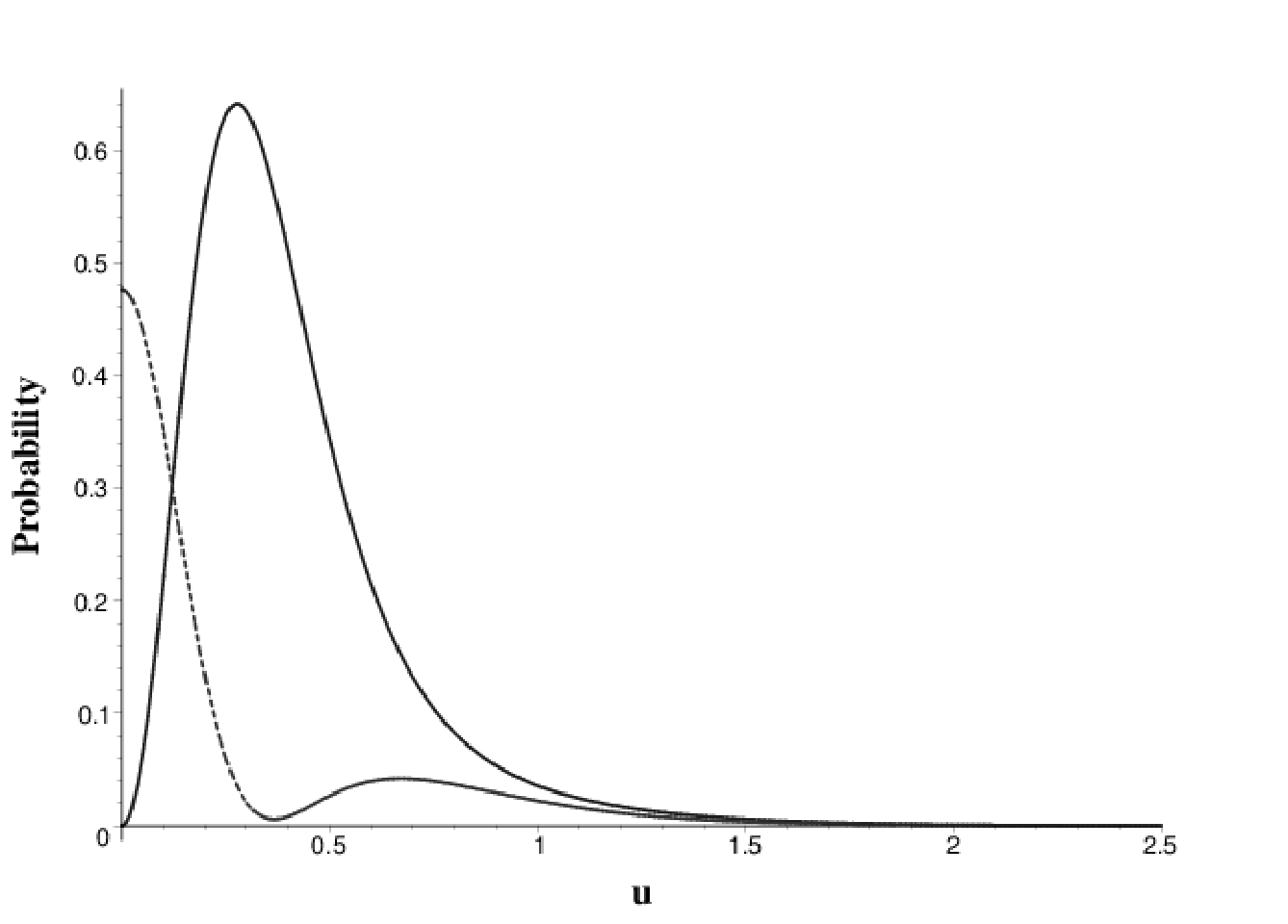}
\caption{$\mathcal{P}_{\gamma \rightarrow e^-e^+}$ as a function of $u$ for $z=0.7$ and $\frac{p}{p\,'}=\frac{1}{2}$. The dashed line represents the case of helicity conservation and the solid line the case when helicity is not conserved.}
\label{f19}
\end{figure}
The first observation that emerge from our graphs Figs.(\ref{f17})-(\ref{f19}) is that the probability of pair production by a single photon is nonvanishing only when the expansion factor was large comparatively with the particle mass. The case $\omega>>m$ corresponds to the early universe conditions. The present days expansion is well approximated by condition $m>>\omega$ and from our graphs for probability and our analytical formulas, we observe that the probabilities of pair production vanish in this limit. Also from our graphs one can see that the probability of production of null mass fermions in the helicity nonconserving case is zero, while in the helicity conserving case it is finite. We note that the zero probability for production of fermions with zero mass in the helicity non-conserving case is the result of conformal invariance. The probability in the helicity nonconserving case is nonvanishing only for nonzero mass, so it is the mass, not the de Sitter background, which breaks the helicity conservation. It is also important to specify that from our graphs of probability it seems that the helicity conservation processes will be dominant as $z$ approaches zero, while for $z$ close to unity, the nonconserving helicity processes will be dominant. This means that the helicity conserving processes will be dominant as long as the momenta of the electron and positron are approximatively equal in modulus $p\simeq p\,'$ and the helicity nonconserving processes will be dominat when one of the momenta is large comparatively with the other momenta, $p>> p\,'$.

In the process of pair production by a single photon the helicity is conserved when for example $\lambda=1,\,\sigma=\sigma'=\frac{1}{2}$. In this case, kinetic parameters can be introduced in the orthogonal local
frame $\{\vec{e}_i\}$ where $\vec{k}=k\vec{e}_3$ and
taking the electron and positron momenta in the plane $(1,3)$, denoting
their spherical coordinates as $\vec{p}=(p,\alpha,0)$ and
${\vec{p}\,}'=(p',\beta,\pi)$ where $\alpha,\, \beta\in(0,\pi)$. Then the probability of pair production in the helicity conserving case, if we explicitly calculate $|\xi_{\frac{1}{2}}^+(\vec{p}\,)
\,\vec{\sigma}\cdot\vec{\varepsilon}_{1}(\vec{k})\,\eta_{\frac{1}{2}}({\vec{p}\,}')|^2$ (\ref{matix}), is proportional with:
\begin{equation}\label{csf}
\mathcal{P}_{\gamma \rightarrow e^-e^+}\sim \cos^2\left(\frac{\alpha}{2}\right)\cos^2\left(\frac{\beta}{2}\right).
\end{equation}

Let us discuss the case of helicity nonconservation. This can happen when $\lambda=1,\,\sigma=\sigma'=-\frac{1}{2}$ and the probability in this case will be proportional with a factor $\mathcal{P}_{\gamma \rightarrow e^-e^+}\sim \sin^2\left(\frac{\alpha}{2}\right)\sin^2\left(\frac{\beta}{2}\right)$.
The main difference appears when we discuss the next two possibilities  $\lambda=1,\,\sigma=\frac{1}{2}\,,\sigma'=-\frac{1}{2}$ and $\lambda=1,\,\sigma=-\frac{1}{2}\,,\sigma'=\frac{1}{2}$ and we must evolve $|\xi_{\pm\frac{1}{2}}^+(\vec{p}\,)
\,\vec{\sigma}\cdot\vec{\varepsilon}_{1}(\vec{k})\,\eta_{\mp\frac{1}{2}}({\vec{p}\,}')|^2$ . In this case the probability is proportional with (\ref{matix}):
\begin{eqnarray}\label{cosi}
\mathcal{P}_{\gamma \rightarrow e^-e^+}\sim\left\{
\begin{array}{cll}
\cos^2\left(\frac{\alpha}{2}\right)\sin^2\left(\frac{\beta}{2}\right)&{\rm for}&\sigma=\frac{1}{2}\,,\sigma'=-\frac{1}{2}\\
\sin^2\left(\frac{\alpha}{2}\right)\cos^2\left(\frac{\beta}{2}\right)&{\rm for}&\sigma=-\frac{1}{2}\,,\sigma'=\frac{1}{2}.
\end{array}\right.\label{ij}
\end{eqnarray}

Let us study now the process in which the electron positron and photon triplet is produced from vacuum. The probability per volume unit for this process is obtained after squaring the amplitude (\ref{as}) and summing after the final helicities $\sigma,\,\sigma\,'$ and $\lambda$:
\begin{eqnarray}\label{pr2}
&&\mathcal{P}_{vac \rightarrow\gamma e^-e^+}=\frac{1}{4}\sum_{\sigma\sigma'\lambda}\frac{|\langle(\vec{p}\,\sigma)_-,(\vec{p}\,'\sigma')_+ ,(\vec{k}\,\lambda)|{\bf S}_1|0
\rangle|^2}{V} =\frac{ e^{2}_{0}}{256\pi}\delta^3(\vec{p}+\vec{p}\,'+\vec{k})\,
|\xi_{\sigma}^+(\vec{p}\,)
\,\vec{\sigma}\cdot\vec{\varepsilon}\,^*_{\lambda}(\vec{k})\,\eta_{\sigma'}({\vec{p}\,}')|^2\nonumber\\
&&\times\left[|n_{\mu}(p,p',k)|^2+|g_{\mu}(p,p',k)|^2+|n_{-\mu}(p,p',k)|^2+|g_{-\mu}(p,p',k)|^2+n_{\mu}(p,p',k)g^*_{\mu}(p,p',k)\right.\nonumber\\
&&\left.+n^*_{\mu}(p,p',k)g_{\mu}(p,p',k)+n^*_{-\mu}(p,p',k)g_{-\mu}(p,p',k)+n_{-\mu}(p,p',k)g^*_{-\mu}(p,p',k)\right.\nonumber\\
&&\left.-{\rm sign}(\sigma\sigma')(n^*_{\mu}(p,p',k)n_{-\mu}(p,p',k)+n^*_{-\mu}(p,p',k)n_{\mu}(p,p',k)+g^*_{-\mu}(p,p',k)g_{\mu}(p,p',k))\right.\nonumber\\
&&\left.-{\rm sign}(\sigma\sigma')(g^*_{\mu}(p,p',k)g_{-\mu}(p,p',k)+n^*_{-\mu}(p,p',k)g_{\mu}(p,p',k)+n_{-\mu}(p,p',k)g^*_{\mu}(p,p',k))\right.\nonumber\\
&&\left.-{\rm sign}(\sigma\sigma')(g^*_{-\mu}(p,p',k)n_{\mu}(p,p',k)+n^*_{\mu}(p,p',k)g_{-\mu}(p,p',k))\right].
\end{eqnarray}
The total probability will be $\mathcal{P}^{tot}_{vac\rightarrow \gamma e^-e^+}=\int \mathcal{P}_{vac\rightarrow \gamma e^-e^+}\,\frac{d^3p}{(2\pi)^{3}}\frac{d^3p\,'}{(2\pi)^{3}}\frac{d^3k}{(2\pi)^{3}}$. As above we do not try to solve the integrals and we analyse the probability (\ref{pr2}) by plotting as a function of the parameter $u=m/\omega$.
\begin{figure}[H]\centering
\includegraphics[scale=0.4]{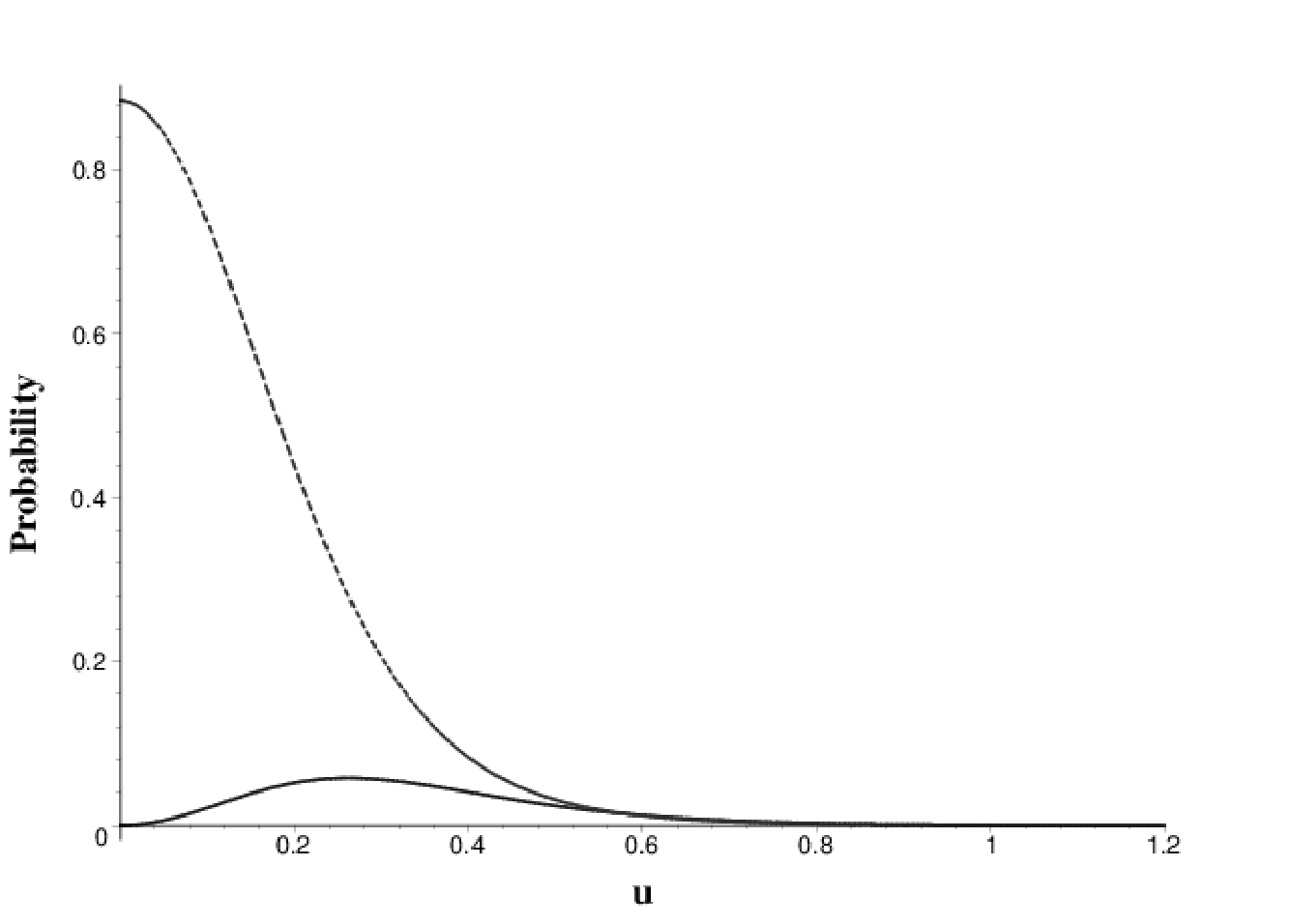}
\caption{$\mathcal{P}_{vac \rightarrow \gamma e^-e^+}$ as a function of $u$ for $z=0.2$ and $\frac{p}{p\,'}=\frac{1}{2}$. The dashed line represents the case of helicity conservation and the solid line the case when helicity is not conserved.}
\label{f20}
\end{figure}

\begin{figure}[H]\centering
\includegraphics[scale=0.4]{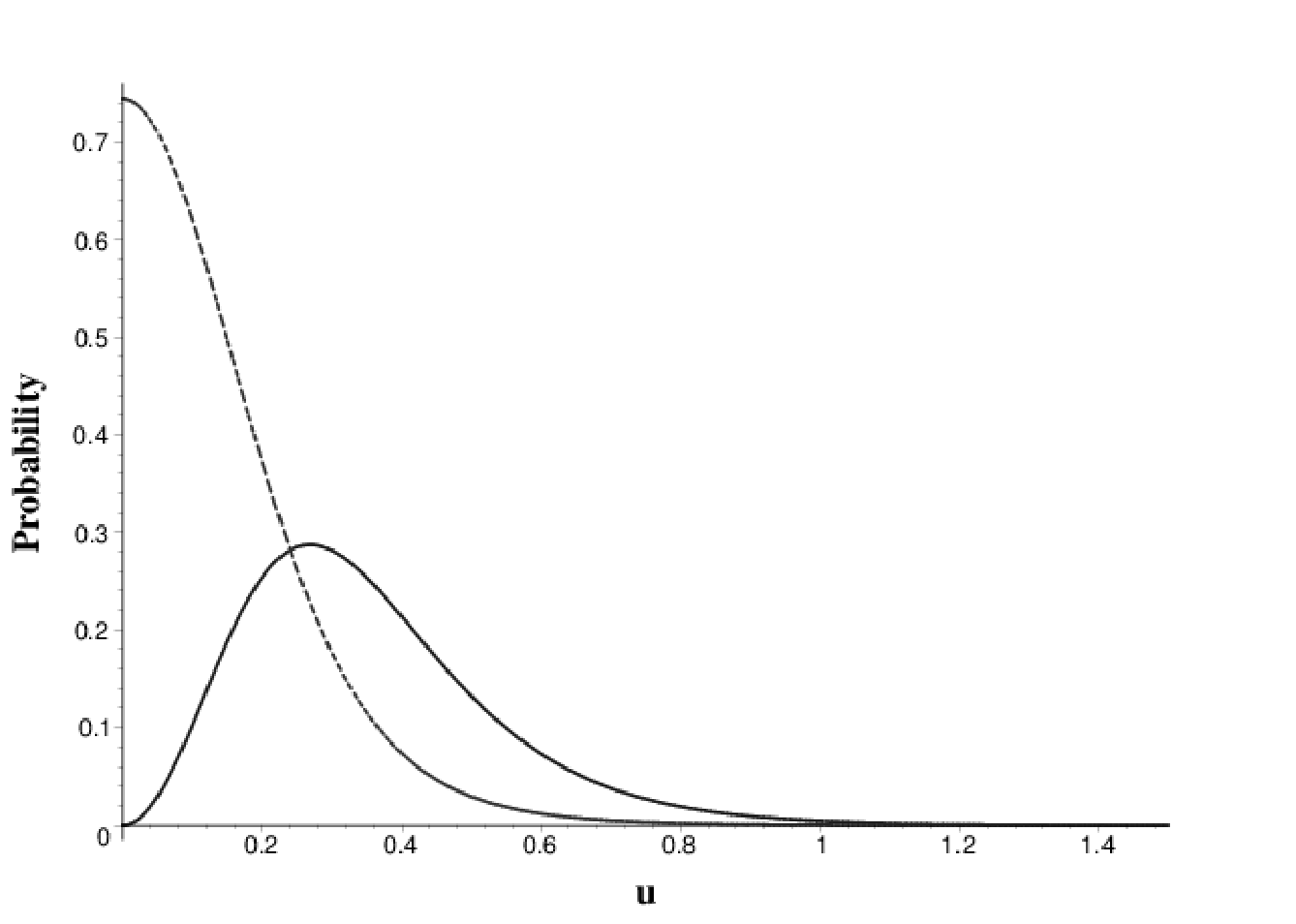}
\caption{$\mathcal{P}_{vac \rightarrow \gamma e^-e^+}$ as a function of $u$ for $z=0.7$ and $\frac{p}{p\,'}=\frac{1}{2}$. The dashed line represents the case of helicity conservation and the solid line the case when helicity is not conserved.}
\label{f21}
\end{figure}

\begin{figure}[H]\centering
\includegraphics[scale=0.4]{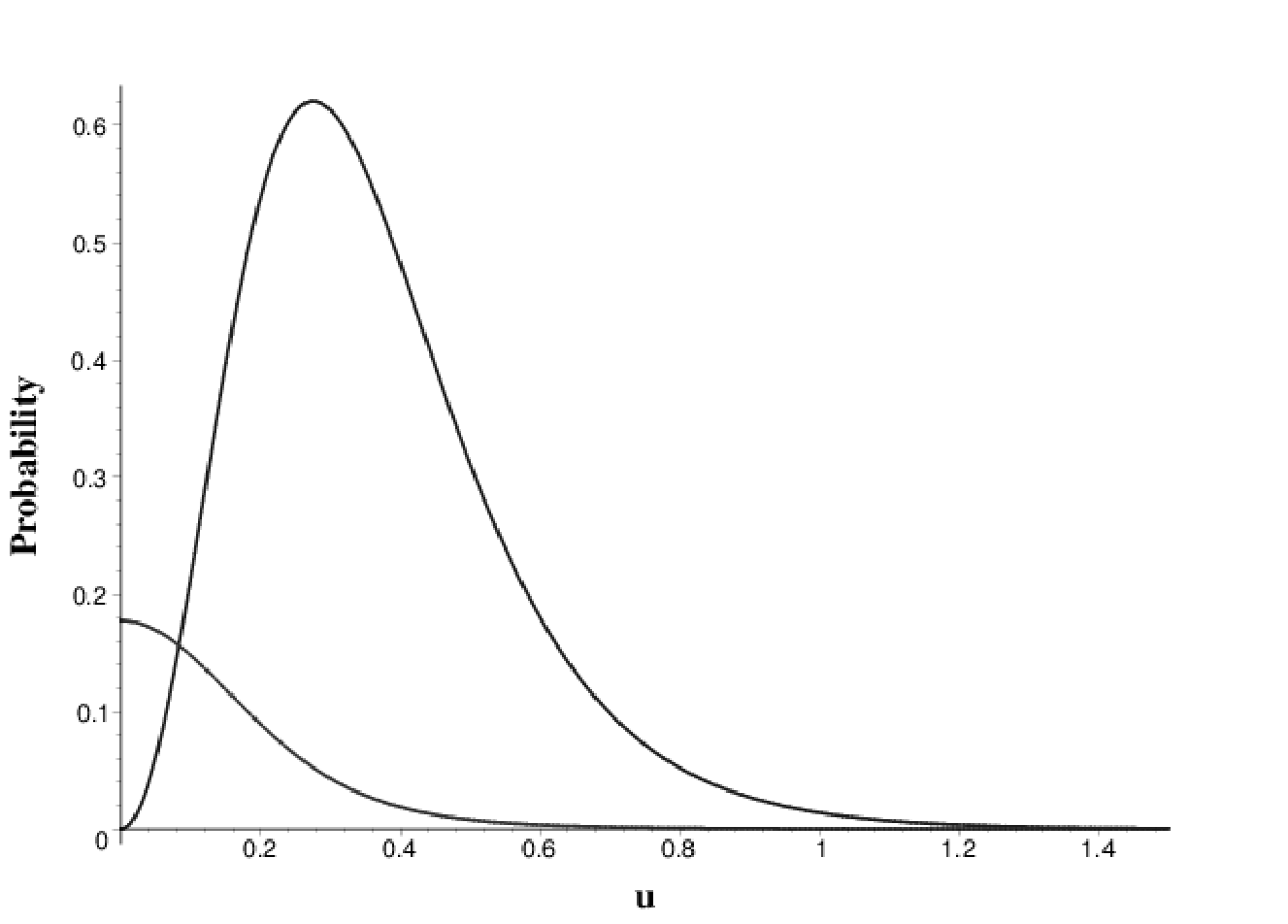}
\caption{$\mathcal{P}_{vac \rightarrow \gamma e^-e^+}$ as a function of $u$ for $z=0.9$ and $\frac{p}{p\,'}=\frac{1}{2}$. The dashed line represents the case of helicity conservation and the solid line the case when helicity is not conserved.}
\label{f22}
\end{figure}
In this case, the processes where helicity is conserved are much more probable when $z$ is close to zero as we observe from Figs.(\ref{f20})-(\ref{f21}). As $z$ is close to unity, the helicity nonconserving processes become dominant (see Fig.(\ref{f22}). The graphs also shows that in both conserving/nonconserving helicity cases the probability is nonvanishing only for $\omega >> m$. Our result concerning the production of particles from vacuum in de Sitter QED is confirmed by the work of L.Parker \cite{WKB}, who establishes that the rate of pair production from vacuum was important only in the early universe. Like in the previous case of pair production by a single photon, there is no production of null mass fermions in the helicity nonconserving case as we can see from Figs.(\ref{f20})-(\ref{f22}).

Helicity conservation in this process is obtained when $\sigma=\sigma\,'=\pm\frac{1}{2},\,\lambda =\mp1$.  Then our probability in the helicity conserving case for $\sigma=\sigma\,'=\frac{1}{2},\,\lambda =-1$ will be proportional with:
$\mathcal{P}_{vac \rightarrow \gamma e^-e^+}\sim \cos^2\left(\frac{\alpha}{2}\right)\cos^2\left(\frac{\beta}{2}\right)$.
For $\sigma=\pm\frac{1}{2},\,\sigma\,'=\mp\frac{1}{2},\,\lambda =-1$ the helicity is not conserved and in this case the probability is proportional with the same factors that appear in Eq. (\ref{cosi}).

Our functions that define the probabilities (\ref{pr1}),(\ref{pr2}), are expressed in terms of hypergeometric functions of the type $F\left(1\pm i\mu,2\mp i\mu;2;\cos^2\frac{\alpha+\beta}{2}\right),F\left(1\pm i\mu,2\mp i\mu;2;\sin^2\frac{\alpha+\beta}{2}\right);$\,\, $\alpha+\beta$ being just the angle between $\vec{p},\,\vec{p\,'}$. Then it is obviously from our analysis from above that our analytical formulas are valid when the momenta of the electron and positron are not parallel . This is because the algebraic argument of hypergeometric functions will became $z=1$, when the momenta are parallel, case in which these functions are divergent.
Even if we can't study the cases when the momenta are parallel, a interesting analysis can be done for small/large angles between the momenta $p,\,p\,'$.
In both processes analysed above, the helicity conservation case is dominant as long as the two momenta $p,\,p\,'$ make between them small angles, close to zero (see (\ref{csf})). Contrary to this, when the angle between momenta is large (close but not equal with $\pi$), the helicity nonconserving processes will be dominant (\ref{cosi}). Hereby we can conclude that only in the helicity nonconserving processes there are chances of separation between electrons and positrons in the early universe. In a helicity conserving process, the electron-positron pair will be emitted at small angles and probably will annihilate each other. When electron-positron pair and a photon are created from vacuum in a helicity conserving process, it is more likely that the fermion pair to annihilate resulting only a photon. To summarise, we can draw the conclusion that the production processes from vacuum, which preserve the helicity conservation law will have the final result only radiation.

At the end of this section we address the problem of Minkowski limit. This corresponds to $\mu=\infty$. Our analytical calculations show that indeed all the functions that define our amplitudes vanish in this limit, $f_{\infty}=g_{\infty}=h_{\infty}=l_{\infty}=n_{\infty}=0$. Also from our graphs Figs.(\ref{f1}-\ref{f16}) we observe that for large $\mu$ the real and imaginary parts of these functions vanish. From here we obtain the Minkowski limit of our amplitudes which give zero. In Minkowski QED these processes are forbidden by the laws of energy-momentum conservation.

\section{Concluding remarks}

We succeeded here to develop the de Sitter QED in Coulomb gauge, as in the flat case, showing how the reduction  formalism and the perturbation procedure allow us to calculate $in-out$ transition amplitudes on the de Sitter expanding universe.  Thus we found that the simplest  effects are those having non-vanishing amplitudes in the first order of perturbations. These are  produced by the classical gravitational field which changes energy with the quantum matter, eliminating the constraints due to the energy-momentum conservation. From our graphical analysis, we obtain that these processes were possible only in the early universe when the expansion factor is larger than the particle mass. Our results confirm the well established results from literature that prove that the rate of pair production was important only in the early universe.

However, we made here only one step to a long way punctuated by many serious difficulties foreshadowed by the analytical forms of our amplitudes which are extremely complicated. We can imagine that the next orders of perturbations as well as the renormalization procedures will give rise to new technical difficulties in working with special functions and solving complicated integrals. The recent studies concerning the regularization of the photon \cite{Wood} and electron-positron \cite{Prot} self-energy diagrams in the second order of perturbations confirm this perspective. Therefore, new mathematical methods are needed for solving these problems if we want to arrive to a strong theory of interacting fields on the de Sitter background, with complete Feynman rules and renormalization in any order.

\appendix

\setcounter{equation}{0} \renewcommand{\theequation}
{A.\arabic{equation}}

\subsection*{Appendix A: Polarization}

\setcounter{equation}{0} \renewcommand{\theequation}
{A.\arabic{equation}}

\subsubsection*{Pauli spinors}

The Pauli spinors  $\xi_{\sigma}$ and $\eta_{\sigma}=i\sigma_2\xi_\sigma^*$ of the {momentum-spin} basis  are defined as $\xi_{\frac{1}{2}}=(1,0)^T$ and $\xi_{-\frac{1}{2}}=(0,1)^T$ for particles and
$\eta_{\frac{1}{2}}=(0,-1)^T$ and $\eta_{-\frac{1}{2}}=(1,0)^T$ for antiparticles \cite{BDR}.  Those of the momentum-helicity basis \cite{CD1}, $\xi_{\sigma}(\vec{p})$ and $\eta_{\sigma}(\vec{p})=i\sigma_2\xi_{\sigma}^*(\vec{p})$, are eigenvectors of the helicity operator,
\begin{equation}\label{heli}
\vec{\sigma}\cdot\vec{p}\,\,\xi_{\sigma}(\vec{p}\,)=2p\,\sigma\,
\xi_{\sigma}(\vec{p}\,) \,, \quad
\vec{\sigma}\cdot\vec{p}\,\,\eta_{\sigma}(\vec{p}\,)=-2p\,\sigma\,
\eta_{\sigma}(\vec{p}\,)\,.
\end{equation}
The polarization is called now helicity since the spin is projected along the momentum direction.  In this basis the particle spinors have the form
\begin{equation}\label{xi}
\xi_{\frac{1}{2}}(\vec{p})=\sqrt{\frac{p_3+p}{2 p}}\left(
\begin{array}{c}
1\\
\frac{p_1+ip_2}{p_3+p}
\end{array} \right)\,,\quad
\xi_{-\frac{1}{2}}(\vec{p})=\sqrt{\frac{p_3+p}{2 p}}\left(
\begin{array}{c}
\frac{-p_1+ip_2}{p_3+p}\\
1
\end{array} \right)\,,
\end{equation}
and satisfy the following properties
\begin{equation}
\sum_{\sigma}\xi_{\sigma}(\vec{p}\,)\,\xi_{\sigma}(\vec{p}\,)^+=1_{2\times
2}\,,\quad
\sum_{\sigma}\sigma\,\xi_{\sigma}(\vec{p}\,)\,\xi_{\sigma}(\vec{p}\,)^+
=\frac{\vec{\sigma}\cdot\vec{p}}{p}
\end{equation}
Similar properties can be deduced  for the anti-particle spinors $\eta_\sigma(\vec{p})$.

\subsubsection*{Polarization vectors}

The polarization of the free Maxwell field is given by the polarization vectors ${\vec\varepsilon}_{\lambda}({\vec k})$ which have c-number components. In the Coulomb gauge these are  orthogonal to the momentum direction, ${\vec k}\cdot{\vec \varepsilon}_{\lambda}({\vec
k})=0$, for any polarization $\lambda=\pm 1$, and satisfy \cite{SW1}
\begin{eqnarray}
{\vec \varepsilon}_{\lambda}({\vec k})\cdot{\vec \varepsilon}_{\lambda'}({\vec
k})^*&=&\delta_{\lambda\lambda'}\,,\label{orto}\\
 \sum_{\lambda}\varepsilon_{\lambda}({\vec k})_i\,\varepsilon_{\lambda}({\vec
k})^*_j&=&\delta_{ij}-\frac{k^i k^j}{k^2}\,.\label{tran}
\end{eqnarray}
Here we consider only  the {\em circular} polarization with
$\vec{\varepsilon}_{\pm 1}(\vec{k})=\frac{1}{\sqrt{2}}(\pm
\vec{e}_1+i\vec{e}_2)$, in a three-dimensional orthogonal local
frame $\{\vec{e}_i\}$ where $\vec{k}=k\vec{e}_3$. These satisfy $\varepsilon_{\lambda}(\vec{k})=-\varepsilon_{\lambda}(-\vec{k})^*$.

\subsubsection*{Polarization matrices}

With these ingredients we can calculate the matrix elements
\begin{equation}
M_{\sigma,\sigma'}(\lambda)=\xi_{\sigma}^+(\vec{p}\,) \,\vec{\sigma}\cdot
\vec{\varepsilon}_{\lambda}(\vec{k})^*\,\eta_{\sigma'}({\vec{p}\,}'),\,\,\,M'_{\sigma,\sigma'}(\lambda)=\xi_{\sigma}^+(\vec{p}\,) \,\vec{\sigma}\cdot
\vec{\varepsilon}_{\lambda}(\vec{k})\,\eta_{\sigma'}({\vec{p}\,}')
\end{equation}
for  $\vec{p}=(p,\alpha,0)$ and ${\vec{p}\,}'=(p',\beta,\pi)$. Using Eqs. (\ref{xi}) and observing that in this case
$\vec{\varepsilon}_{\pm 1}(\vec{k})^*=\frac{1}{\sqrt{2}}(\pm
\vec{e}_1-i\vec{e}_2)$ we obtain the polarization matrices
\begin{eqnarray}\label{matix}
M(1)&=&\sqrt{2}\,\left(
\begin{array}{cc}
-\sin\frac{\alpha}{2}\sin\frac{\beta}{2}&\sin\frac{\alpha}{2}\cos\frac{\beta}{2}\\
-\cos\frac{\alpha}{2}\sin\frac{\beta}{2}&\cos\frac{\alpha}{2}\cos\frac{\beta}{2}
\end{array}\right)=-M'(-1)\,,\\
M(-1)&=&\sqrt{2}\,\left(
\begin{array}{cc}
\cos\frac{\alpha}{2}\cos\frac{\beta}{2}&\cos\frac{\alpha}{2}\sin\frac{\beta}{2}\\
-\sin\frac{\alpha}{2}\cos\frac{\beta}{2}&-\sin\frac{\alpha}{2}\sin\frac{\beta}{2}
\end{array}\right)=-M'(1)\,,
\end{eqnarray}
for $\lambda =1$ and $\lambda =-1$ respectively.

\subsection*{Appendix B: Bessel functions}

\setcounter{equation}{0} \renewcommand{\theequation}
{B.\arabic{equation}}

The Bessel functions of index $\frac{1}{2}$ are elementary
functions,
\begin{equation}\label{Ap1}
K_{\frac{1}{2}}(z)=\sqrt{\frac{\pi}{2z}}\,e^{-z}\,, \quad
H_{\frac{1}{2}}^{(1)}(z)=- i \sqrt{\frac{2}{\pi z}}\,e^{ i z}\,,
\quad H_{\frac{1}{2}}^{(2)}(z)= i \sqrt{\frac{2}{\pi z}}\,e^{- i
z}\,.
\end{equation}
The Hankel functions  we use here
\begin{eqnarray}
H^{(1)}_{\nu_\pm}(z)&=&\frac{e^{\pm\pi\mu}J_{\nu_\pm}(z)
-iJ_{-\nu_\pm}(z)}{\cosh(\pi\mu)}\label{h1}\,,\\
H^{(2)}_{\nu_\pm}(z)&=&\frac{e^{\mp\pi\mu}J_{\nu_\pm}(z)
+iJ_{-\nu_\pm}(z)}{\cosh(\pi\mu)}\,,\label{h2}
\end{eqnarray}
have the limits,
\begin{equation}\label{limits}
\lim_{{x}\to 0}x^{\nu} H^{(1)}_{\nu}(\alpha x)=-\lim_{{x}\to
0}x^{\nu} H^{(2)}_{\nu}( \alpha
x)=\frac{1}{i\pi}\left(\frac{2}{\alpha}\right)^{\nu}\Gamma(\nu)\,,
\end{equation}
that hold since $\Re \nu_{\pm} =\frac{1}{2} >0$ \cite{GR}.

Eq. (\ref{Ap1}a) helps us to evaluate the integrals (\ref{int0}) by
replacing the exponential function in Eq. (\ref{int0}). We obtain thus two types of integrals which can be put in analytical forms. The first integral  \cite{GR},
\begin{equation}\label{int1}
\int_0^{\infty}dx\,x^{\frac{3}{2}}\,K_{\frac{1}{2}}(cx)J_{\nu}(ax)J_{\nu}(bx)
=-\frac{1}{\sqrt{2\pi}}\frac{c^{\frac{1}{2}}}{(ab)^{\frac{3}{2}}}
\frac{d}{du}Q_{\nu-\frac{1}{2}}(u)\,,
\end{equation}
depends on the new variable $u$ which obeys $2abu=a^2+b^2+c^2$. The
second integral under consideration \cite{GR},
\begin{eqnarray}\label{int2}
&&\int_0^{\infty}dx\,x^{\frac{3}{2}}\,K_{\frac{1}{2}}(cx)J_{\nu}(ax)J_{-\nu}(bx)\nonumber\\
&&~~~~~~~~~~~~~=\frac{\sin
\pi\nu}{\sqrt{2\pi}\,\nu}\,c^{-\frac{5}{2}}\left(\frac{a}{b}\right)^{\nu}
F_4\left(\frac{3}{2},1,1+\nu,1-\nu;-\frac{a^2}{c^2},-\frac{b^2}{c^2}\right) \,,
\end{eqnarray}
is solved in terms of Appell hypergeometric functions $F_4$ depending on double
arguments. Both these integrals are convergent for $\Re(c)>0$.  Therefore, in order to evaluate Eqs. (\ref{Apm}),  (\ref{Cpm}) and (\ref{Bpm}) we must
replace $c\to \epsilon-iq$, introducing thus the usual small $\epsilon>0$ which
finally tends to zero.

\subsection*{Appendix C: Legendre functions}

\setcounter{equation}{0} \renewcommand{\theequation}
{C.\arabic{equation}}

The Legendre function of the second kind \cite{GR} can be written as
\begin{equation}
Q_\nu(z\pm i0)=\frac{\pi}{2\sin\pi\nu}\left[\,e^{\mp
i\pi\nu}P_\nu(z)-P_\nu(-z)\right]
\end{equation}
for $-1<z<1$. The Legendre functions of the first kind,
\begin{equation}
P_\nu(z)=F\left(-\nu,1+\nu;1;\frac{1-z}{2}\right)\,,
\end{equation}
are analytic in this domain being represented by the usual Gauss hypergeometric
function $F\equiv\, _2F_1$. Hereby we obtain the formula
\begin{eqnarray}
\frac{d}{dz}\,Q_\nu(z\pm i0)&=&\frac{\pi\nu(\nu+1)}{4\sin\pi\nu}\left[e^{\mp
i\pi\nu}F\left(1-\nu,2+\nu;2;
\frac{1-z}{2}\right)\right.\nonumber\\
&&\left.\hspace{25mm}+\,F\left(1-\nu,2+\nu;2;\frac{1+z}{2}\right)\right]
\end{eqnarray}
which helps us to calculate the functions $A_\pm$ (for $\nu=\pm i\mu$) and
$C_\pm$ (taking $\nu=\mp i\mu-1$).

\end{document}